\documentclass[referee,a4paper,12pt,traditabstract]{swsc_MH} 

\usepackage{graphicx}
\usepackage{txfonts}
\usepackage{subfigure}
\usepackage{epstopdf}
\usepackage[backref]{hyperref}
\usepackage{lineno}
\usepackage{natbib}
\usepackage[normalem]{ulem}
\hypersetup{colorlinks=true,citecolor=cyan,urlcolor=cyan,linkcolor=blue}
\begin{document}
   \title{Identification of photospheric activity features from SOHO/MDI data using the ASAP tool}
   \titlerunning{Identification of activity features from SOHO/MDI}
   \authorrunning{Ashamari et al.,}
   \author{Omar Ashamari
          \inst{1}
          \and
          Rami Qahwaji          
          \inst{1}
          \and
          Stan Ipson
          \inst{1}          
          \and                                                        
          Micha Sch\"oll
          \inst{2}      
          \and         
         Omar Nibouche
          \inst{3}
          \and
          Margit Haberreiter
          \inst{4}  
       } 

   \institute{School of Electrical Engineering and Computer Science, University of Bradford, Bradford, UK   	
          \email{\href{mailto:O.W.A.Ashamari@bradford.ac.uk}{O.W.A.Ashamari@bradford.ac.uk}}
         \and
         LPC2E/CNRS, 3A Av. de la Recherche Scientifique, 45071 Orl\'eans, France
         \and         
         School of Computing and Mathematics, University of Ulster, BT37 0QB, Newtownabbey, UK
         \and         
         Physikalisch-Meteorologisches Observatorium and World Radiation Center, Dorfstrasse 33, CH-7260 Davos Dorf, Switzerland   
         }
            
   \date{Received October 23, 2013; accepted May 05, 2015}


  \abstract
 	{The variation of solar irradiance is one of the natural forcing mechanisms of the terrestrial climate. Hence, the time-dependent solar irradiance is an important input parameter for climate modelling. The solar surface magnetic field is a powerful proxy for solar irradiance reconstruction. The analyses of data obtained with the Michelson Doppler Imager (MDI) on board the SOHO mission are therefore useful for the identification of solar surface magnetic features to be used in solar irradiance reconstruction models. However, there is still a need for automated technologies that would enable the identification of solar activity features from large databases. To achieve this we present a series of enhanced segmentation algorithms developed to detect and calculate the area coverages of specific magnetic features from MDI intensitygrams and magnetograms. These algorithms are part of the Automated Solar Activity Prediction (ASAP) tool. 
 		
	The segmentation algorithms allow us to identify the areas on the solar disk covered by magnetic elements inside and outside boundaries of active regions. Depending on their contrast properties, magnetic features within an active region boundary are classified as sunspot umbra and penumbra, or faculae. Outside an active region boundary magnetic elements are identified as network. We present the detailed steps involved in the segmentation process and provide the area coverages of the segmented MDI intensitygrams and magnetograms. The feature segmentation has been carried out on daily intensitygrams and magnetograms from April 21, 1996 to April 11, 2011. This offers an exciting opportunity to undertake further investigations that benefit from solar features segmentations, such as solar irradiance reconstruction, which we plan to investigate in the future.
	}  
   \keywords{Photosphere -- 
   			Sun --
   			Solar cycle -- 
            Spectral irradiance
            }
	
   \maketitle
	
\section{Introduction}
Solar radiation is the Earth's most important energy source. It is not constant, but varies over all observed timescales. As such the incoming solar irradiance is a very significant natural driver for the Earth's climate. For recent reviews see \cite{Solanki2013ARAA}, \cite{Ermolli2013ACP} and \citet[][Space Sci. Rev.]{Yeo2014SSRv}. Depending on its wavelength, the incoming radiation is absorbed in different layers of the Earth's atmosphere and deposits its energy accordingly. For a detailed study of the influence of solar irradiance, including the Solar Spectral Irradiance (SSI) and the Total Solar Irradiance (TSI), on the Earth's climate system it is however crucial to know accurately its variations over different time scales. Various methods of reconstruction of both the SSI and the TSI have been presented in the literature. These methods can be categorised into proxy-based and semi-empirical models.

The proxy-based models study to what extent specific solar proxies, such as sunspot number, Mg II index, F10.7 radio flux, {abundance of} cosmogenic isotopes, spectral lines or specific wavelength bands in the observed solar spectra or solar intensitygrams, can be used to describe solar irradiance variability. The NRLSSI model \citep[see e.g., ][]{Lean2011JGR_EUV} is based on the {sunspot number,} Mg II and F10.7 indices, and covers the full solar spectrum from the EUV to the IR. \cite{Thuillier2012} employ the Mg II index along with neutron monitoring observations, with the model named MGNM to model variations in the UV spectral range, while \cite{Morril2011_UV} use the Mg II index and the Ca II K activity derived from ground based observations along with high-resolution spectra taken by the Solar Ultraviolet Spectral Irradiance Monitor \cite[SUSIM,][]{Brueckner1993} on the Upper Atmosphere Research Satellite (UARS) to model the variations in the UV part of the spectrum. 
\cite{DudokDeWit2014} employ a set of radio observations to empirically model the SSI from the EUV to the visible part of the spectral range. \cite{Preminger2011} analyse the effect of the photometric sum of the solar features identified from images obtained at the San Fernando Observatory \citep[SFO,][]{Chapman1997} in the red and blue continuum {to model the variation in TSI}. For the understanding of the UV spectral range in particular, \cite{Lean1998} and \cite{Worden1998,Worden2001} analysed Ca II K spectroheliograms to derive information on the evolution of solar activtiy features used in their empirical models.


The semi-empirical models are based on the fact that various distinct activity features on the solar disk emit characteristic spectra. This involves determining the time-dependent area coverage of features such as sunspots, faculae and network for any given point in time. The contrast of the various features varies with wavelength and distance from disk centre. The area coverage of the activity features {is} determined from solar images and/or magnetograms. Finally, weighting the intensity spectra with the position-dependent relative areas covered by the features yields the temporal variation of the SSI \citep[for an overview see e.g.,][]{Ermolli2013ACP}. Different groups use different spectral synthesis codes, solar images and magnetogram data. 

Full-disk observations carried out with the Precision Solar Photometric Telescope in Rome \citep[PSPT,][]{Ermolli2010AA} are employed in the Osservatorio Astronomico di Roma (OAR) model of solar irradiance variations \citep{Domingo2009,Ermolli2011}. More recently, \cite{Ermolli2013ACP} used, along with the PSPT images, spectra calculated with the RH radiative transfer code \citep{Uiten2002} to reconstruct TSI and the SSI from the UV to the IR wavelength range. Moreover, \cite{Fontenla2011} reconstruct TSI and the SSI from the EUV to the IR using the Solar Radiation Physical Modeling tools \citep[SRPM,][]{Fontenla2009ApJ,Fontenla2011} along with data from PSPT \citep{PSPT}, also operating at the Mauna Loa Solar Observatory. \cite{Haberreiter2014} use the Solar Modeling code \citep[SOLMOD,][]{Haberreiter2011, Haberreiter2012} and images obtained with SOHO/EIT for the identification of activity features with the SPoCA tool \citep{Verbeeck2014} to reconstruct the solar EUV spectral range. 

Synthetic spectra calculated in non-Local Thermodynamic Equilibrium (non-LTE) with the COde for Solar Irradiance \citep[COSI,][]{Haberreiter2008a, Shapiro2010} has also been used for semi-empirical modelling of solar irradiance \citep{Haberreiter2005ASpR,Shapiro2011AA}. With the analysis of MDI magnetograms presented here we aim at improving their UV reconstruction.

Recently \cite{Yeo2014AA} used a series of full-disk observations from the Kitt Peak Vacuum Telescope \cite[KPVT,][]{KPVT}, SOHO/MDI and the Helioseismic and Magnetic Imager on board the Solar Dynamics Observatory \cite[SDO/HMI,][]{SDO} to reconstruct TSI and SSI from the FUV to the IR, with the SATIRE-S model for the period of 1974-2013 \citep[see e.g.,][and references therein]{Krivova2003,Unruh2008,Ball2011,Yeo2013}. In the following we summarise the concept of the SATIRE reconstruction model in more detail as it is in some respect similar to the new approach presented here.
	
Sunspots are identified as groups of magnetic pixels that have continuum intensities below a certain threshold value. All other magnetic pixels on the solar disk are identified as faculae and network. In the irradiance reconstruction these pixels are assigned a specific weighting, i.e. filling factor. It has a value between 0 and 1, depending on the pixel's magnetic field strength and is determined via linear interpolation between 0 G and the saturation value ($B_{\rm sat}$) of the faculae and network pixels. Finally, the intensity spectra for the different features and positions on the disk are calculated with the LTE spectral synthesis code ATLAS9 \citep{Kurucz1991}. For the faculae and network elements the same intensity spectra are used.

It is clear that each of the groups that identify the area coverage of solar activity features from various instruments use slightly different definitions and identifications for these features. {We emphasise that there are different definitions for some photospheric solar structures, and there is no universally accepted unique definition for these structures \citep{HarveyWhite1999}.} Sunspots, for example, are typically straight forward to identify as they appear dark in photospheric intensitygrams. The {differences arise} for faculae and network {definitions and} identification {techniques. For instance,} using PSPT images, faculae and network are identified through the position-dependent contrast of the narrow-band Ca II K intensities. On the number of features identified, several discrete thresholds might be applied. Therefore, unless the same {feature identification and consequently} threshold {technique} is applied by different authors, a direct comparison of the segmentation results is difficult. 

{Here we present yet another definition for the detected solar photospheric structures taking into account the locations of magnetic pixels with respect to active regions.} We also use continuum intensitygrams and magnetograms, however the idea is that, besides the information of the contrast and the photospheric magnetic field strength of the features, their {proximity} to an active region is also taken into account. In particular, we investigate to what extent faculae and network can be distinguished through their {proximity} to a sunspot group.

The present work is the first of a series of papers that aims toward creating a new approach for the identification of solar features from SOHO/MDI intensitygrams and magnetograms, for the purpose of semi-empirical irradiance reconstruction. The motivation for this is that in the near future we aim to {incorporate it into} an automated system that can operate in a real-time mode to process low-latency data from space missions. As a proof of concept we start our investigations on SOHO/MDI data, and in the near future we plan to investigate SDO/HMI data and adapt our methods accordingly. One particular outcome of the current work is the provision of area coverages of different solar features, which will be used at a later stage for irradiance reconstruction, and would also be available to the community to carry out any kind of analysis for a better understanding of these features. In this paper we discuss and present the detection methodology of solar features from SOHO/MDI intensitygrams and magnetograms and their area coverage. For this purpose, we use an extended version of the Automated Solar Activity Prediction tool \citep[ASAP,][]{ColakQahwaji2008,ColakQahwaji2009}, which was used in previous research on the detection of sunspots and active regions. We detect magnetically active solar features in magnetograms and locally dark solar features in intensitygrams and provide active regions, sunspot umbra and penumbra, faculae, and network detections as well as the quiet Sun {at daily intervals for the available MDI intensitygrams and magnetograms}. 

We provide binary segmentation maps along with the solar area coverages of the features as a function of position on the solar disk. In an upcoming paper we will assign intensity spectra to these activity features and reconstruct the SSI for the full MDI time series.

The following parts of the paper are organised as follows: The SOHO/MDI data used in this study and the pre-processing applied are briefly described in Section\,\ref{sec:data}. The statistics of the intensity and magnetogram signals from the full set of the MDI data are discussed in Section\,\ref{sec:dataanalysis}. The methodologies adopted for the detection of features in solar images are discussed in Section\,\ref{Sec:MDISeg}, the results obtained and their analysis, including a simple model of TSI based on the results obtained here, are presented in Section\,\ref{sec:results}, while conclusions are drawn in Section\,\ref{sec:concl}.

\section{MDI Data}\label{sec:data}
The Solar Oscillations Investigation/Michelson Doppler Imager \citep[SOI/MDI][]{MDI} on board of the Solar and Heliospheric Observatory (SOHO) satellite took sets of high quality full-disk images. The images were taken using five narrow-band filters through the Ni I 676.8 nm absorption spectral line, and were used to produce, amongst others, the intensitygrams and magnetograms \footnote{Full-disk 1024$\times$1024 pixel intensitygrams and line-of-sight magnetograms data sets are both available to download from the SOHO and JSOC archive: http://soho.nascom.nasa.gov; http://jsoc.stanford.edu}. 

In this work, we use level 2 intensitygrams, which have been corrected for limb-darkening, observed with a 6-hour cadence and level 1.8 5-min magnetograms, observed with a 96-minute cadence. A total of 90,463 FITS data files were downloaded, which included 17,347 intensitygrams and 73,116 magnetograms. The downloaded data cover a period from April 21, 1996, when the first magnetogram was recorded, to April 11, 2011, when the MDI observations were terminated. This period covers solar cycle 23 and the beginning of solar cycle 24, and includes data from two solar minima and one solar maxima periods. The downloaded data set, however, includes a significant number of corrupted intensitygrams and magnetograms, which if left included, would cause miss detection and classification of solar features. Therefore, care was taken to identify and exclude corrupted files, which have poor quality and missing values.

For the purpose of the work described here, one pair of intensitygram and magnetogram is selected for each day over the complete MDI data set. These were determined, by selecting the intensitygram observed closest to 12:00 UT each day. Then the magnetogram closest in time to the intensitygram was chosen. As part of the selection process, the time difference between the selected intensitygram and magnetogram pair was not allowed to exceed three hours. Each intensitygram was then re-mapped taking account of differential solar rotation, using the approach discussed in \cite{Tang1981}, to correspond to the observing time of the associated magnetogram. Through this process, 2,490 days with joint intensitygram-magnetogram pairs remained and were used for the subsequent analysis that is discussed here.

\section{Data Analysis}\label{sec:dataanalysis}
 In this work, we take advantage of the ASAP tool for initial photospheric feature detection and grouping. ASAP was designed to operate in near real-time and so used SOHO/MDI intensitygrams and magnetograms, provided in Graphics Interchange Format (GIF) 8-bit format. However, the downloaded data files used in the current work are in the Flexible Image Transport System (FITS) format, containing data stored as signed 32-bit format as well as detailed header information. Therefore, the conversion of the FITS data from 32-bit signed to unsigned 8-bit greyscale images is necessary to enable the data visualisation and form images that are similar in nature to the GIF ones provided in SOHO archives. 
To implement the data type conversion, the solar disk minimum and maximum values of the investigated intensitygrams and magnetograms were determined. We found the full range of values are between 0 and 1.5 for the intensitygrams, and between -3,500 and 3,500 Gauss for the magnetograms. These values were therefore used in linear operations to scale the signed 32-bit data from the FITS files to unsigned 8-bit data, which were then stored as images. 

Subsequently, we found and recorded the solar disk mean, standard deviation, minimum, and maximum values for each of the investigated intensitygrams and magnetograms. This is a very useful step for analysing and understanding the nature of the investigated data set. The collected values are therefore analysed for the purpose of finding parameters that enable the efficient and constant detection and enhanced visualisation of the photospheric features. The collected values and their analysis are presented and discussed in the following subsections. 

\subsection{Intensitygram Analysis}
The mean ($\left\langle I \right\rangle$), standard deviation ($\sigma _{\rm I}$), minimum ($I_{\rm min}$) and maximum ($I_{\rm max}$) values of the on-disk pixels for the intensitygrams are recorded and analysed in order to find the most appropriate parameters and values that enable the detection and visualisation of sunspots, pores and quiet Sun. These values are summarised in Table\,\ref{T:stats} and discussed in the following subsections. We report the values of these parameters as relative 8-bit pixel values, therefore we do not assign units to them.

\begin{table}[h!]
	\caption{Summary of the statistical analysis of the intensitygrams. }
	\begin{center}
		{\footnotesize 
			\begin{tabular}{lcl}
				\hline
				\noalign{\smallskip}
				\noalign{\smallskip}
				Parameter & 8-bit Value & Description \\
				\noalign{\smallskip}
				\noalign{\smallskip}
				\hline
				\noalign{\smallskip}
				\noalign{\smallskip}
				
				$\left\langle \left\langle I \right\rangle \right\rangle$,  & 169.5 &  The mean and median value of $\left\langle I \right\rangle$ during\\				
				$\left\langle \left\langle I \right\rangle \right\rangle_{\rm med}$ &  & the solar cycle. \\
				\noalign{\smallskip}				
				$\left\langle \sigma _{\rm I,SMin}\right\rangle,$ & 1.5 & The mean and median value of a set of $\sigma _{\rm I}$ during\\
				$\left\langle \sigma _{\rm I,SMin}\right\rangle _{\rm med}$ &  & the Solar Minimum (SMin) period.\\								
								
				\noalign{\smallskip}
				\hline
			\end{tabular}
			\label{T:stats}}
	\end{center}
\end{table}

\subsubsection{Intensitygram Statistics}
\begin{figure*}[ht!]
	\centering
	\includegraphics[width=0.80\textwidth]{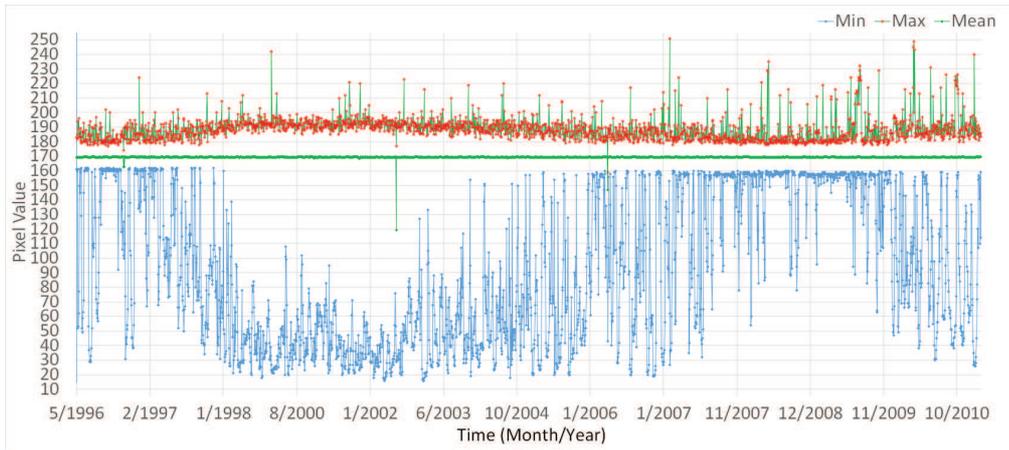}
	\caption{\label{fig:intensitygram}
		$\left\langle I\right\rangle$ (green), $I_{\rm min}$ (blue) and  $I_{\rm max}$ (red) of the solar disk for the full set of 8-bit intensitygrams. }
\end{figure*}

We collected $\left\langle I\right\rangle$, $I_{\rm min}$ and  $I_{\rm max}$ values within the solar disk for the full set of the analysed intensitygrams. Figure\,\ref{fig:intensitygram} shows plots of $\left\langle I\right\rangle$ (green), $I_{\rm min}$ (blue) and  $I_{\rm max}$ (red) for the on-disk pixels of the intensitygrams over the full MDI time series. We found $\left\langle I\right\rangle$ to be fairly steady with average about 169.5. We found the $I_{\rm min}$ values to be higher on average during the solar minimum period in comparison with during the solar maximum period. The collected $I_{\rm min}$ values can represent the intensities of sunspots, pores, or the quiet Sun. Therefore we see the $I_{\rm min}$ values drop to its lowest levels during the solar maximum, when bigger sunspots with lower intensities are common. We found $I_{\rm max}$ values to increase slightly during the solar maximum, when compared to during the solar minimum period. The $I_{\rm max}$ values can represent the intensities of the quiet Sun or bright regions. It is also worth noting that extreme values are often found to correspond to single erroneous pixels. 

\begin{figure*}[ht!]
	\centering
	\includegraphics[width=0.80\textwidth]{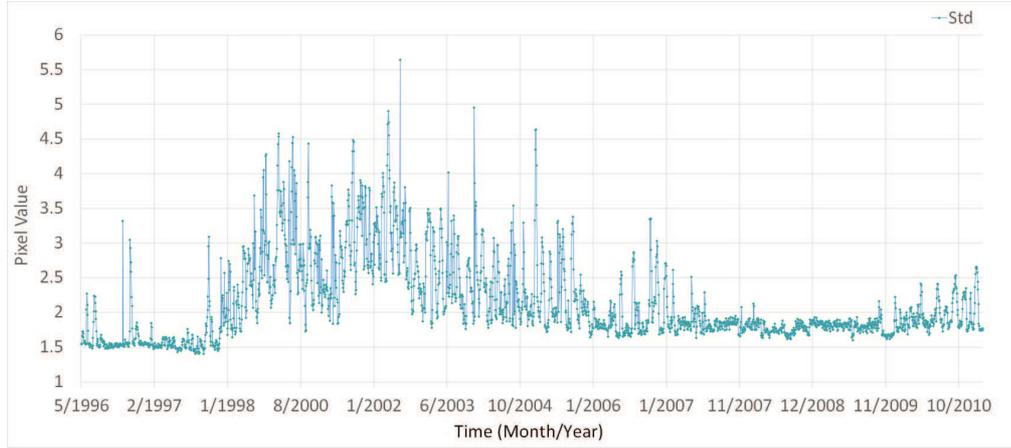}
	\caption{\label{fig:std}
		Standard deviation, $\sigma_{\rm I}$, of the solar disk for the full set of 8-bit intensitygrams.}
\end{figure*}

\begin{figure*}[t!]
	\centering
	\includegraphics[width=0.80\textwidth]{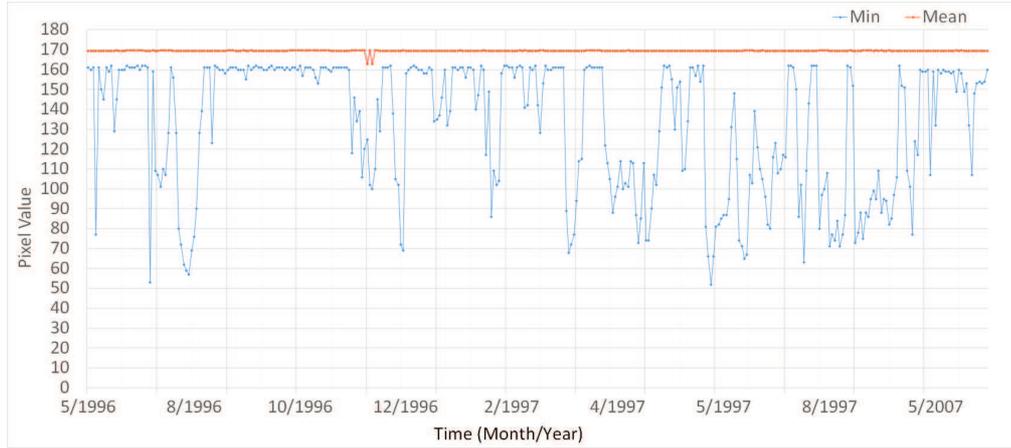}
	\caption{\label{fig:min}
		$I_{\rm min}$ (blue) and $\left\langle I_{\rm SMin}\right\rangle$ (red) of 350 intensitygrams, with the lowest $\sigma_{\rm I}$, collected from the solar minimum periods.}
\end{figure*}

Moreover, we determined the standard deviation $\sigma_{\rm I}$ for on-disk pixels of individual intensitygrams for the full set. The collected $\sigma_{\rm I}$  are shown plotted in Figure\,\ref{fig:std}. As expected, we found $\sigma_{\rm I}$ to be lowest during the solar minimum periods and to increase gradually towards the solar maximum period or with the presence of solar activity. Thus, we find $\sigma_{\rm I}$ to be a good indicator for selecting intensitygrams at times with low solar activity. Figure\,\ref{fig:min} shows the statistics of the intensitygrams selected during quiet time. 

\subsubsection{Sunspot Intensities}\label{Sec:SSInt}
We collected and analysed 350 intensitygrams from the solar minimum periods for the purpose of obtaining an initial sunspot detection threshold ${Th_{\rm ss}}$. For each image, we analysed the difference between $I_{\rm min}$ and $\left\langle I\right\rangle$. The collected data are shown plotted in Figure\,\ref{fig:min}. As expected intensitygrams with the smallest differences between $I_{\rm min}$ and $\left\langle I\right\rangle$ have no sunspots or pores. In cases where the difference between $I_{\rm min}$ and $\left\langle I\right\rangle$ is larger, it turns out that the corresponding intensitygrams contain pores or small sunspot groups. We experimented with the collected data to empirically define the sunspot threshold, by linearly increasing the factor $f_{\rm ss}$ subtracted from $\left\langle I\right\rangle$ of the investigated images, and found that $f_{\rm ss}=15$ in the 8-bit range of values provides a reasonable, by visual inspection, threshold for the sunspot detection. Thus we adopt the following equation for calculating the initial $Th_{\rm ss}$ from the intensitygrams:

\begin{equation}
Th_{\rm ss}=\left\langle I \right\rangle - f_{\rm ss}. \label{E:stats}
\end{equation}

\begin{figure*}[t!]
	\centering
	\includegraphics[width=0.29\textwidth]{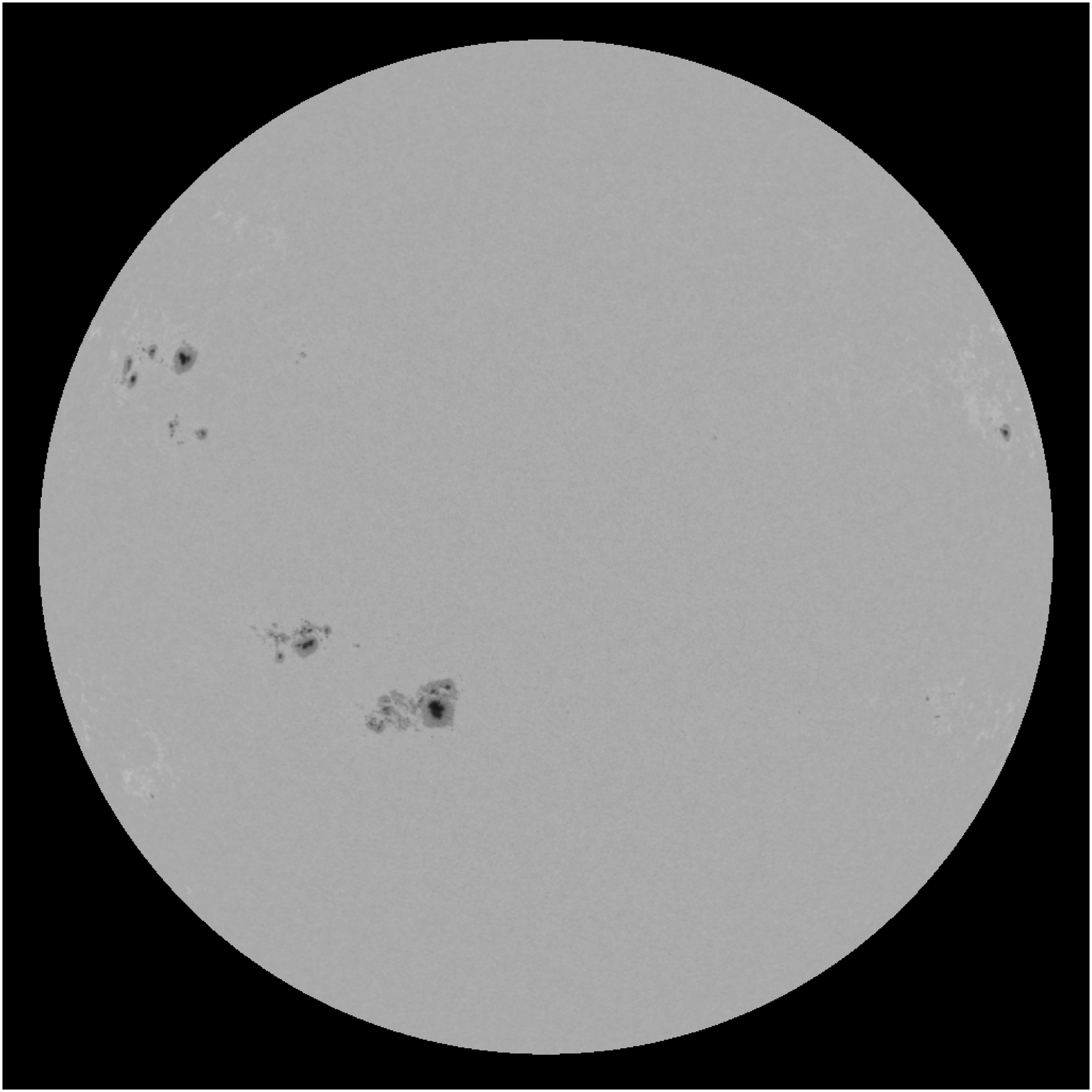}
	\includegraphics[width=0.29\textwidth]{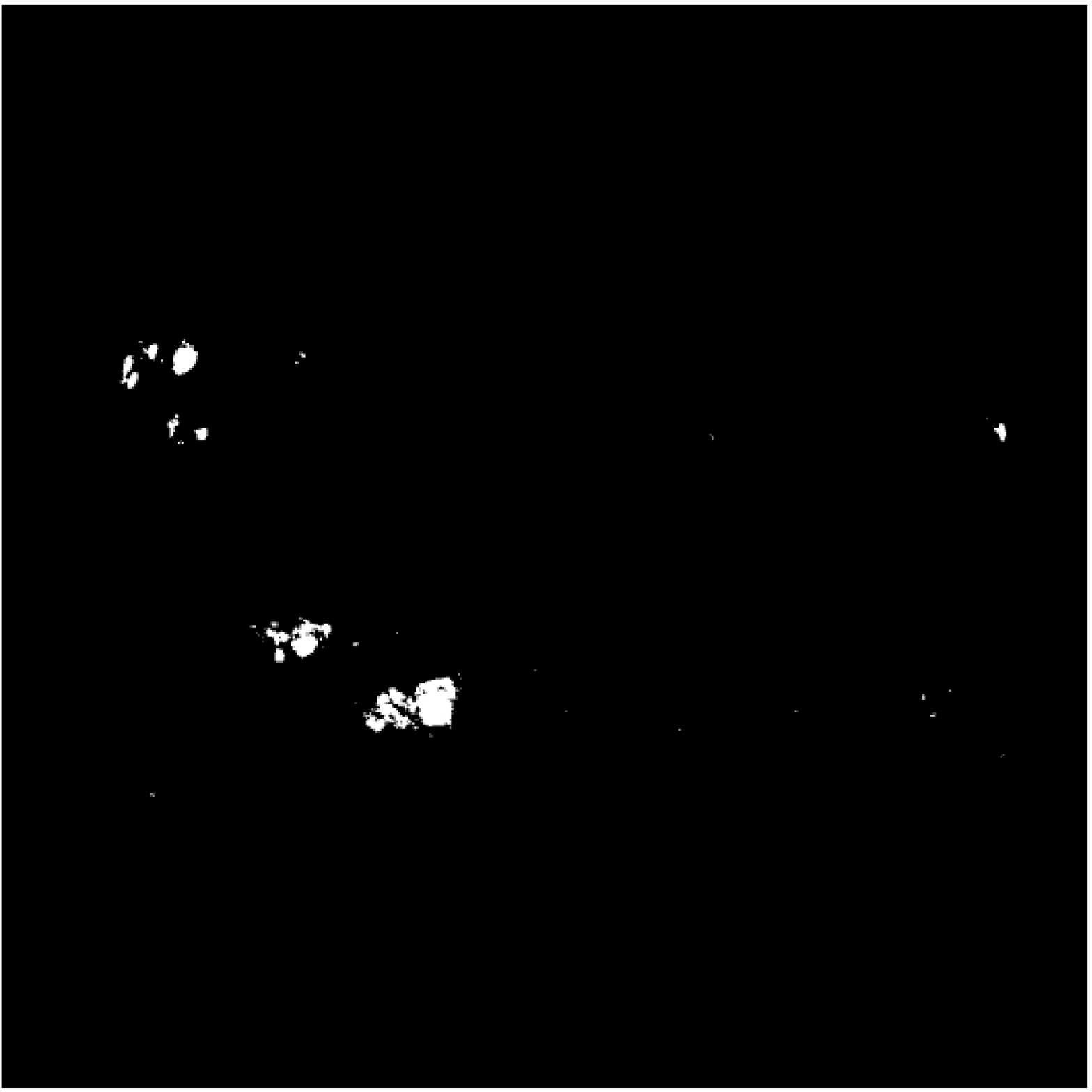}
	\includegraphics[width=0.29\textwidth]{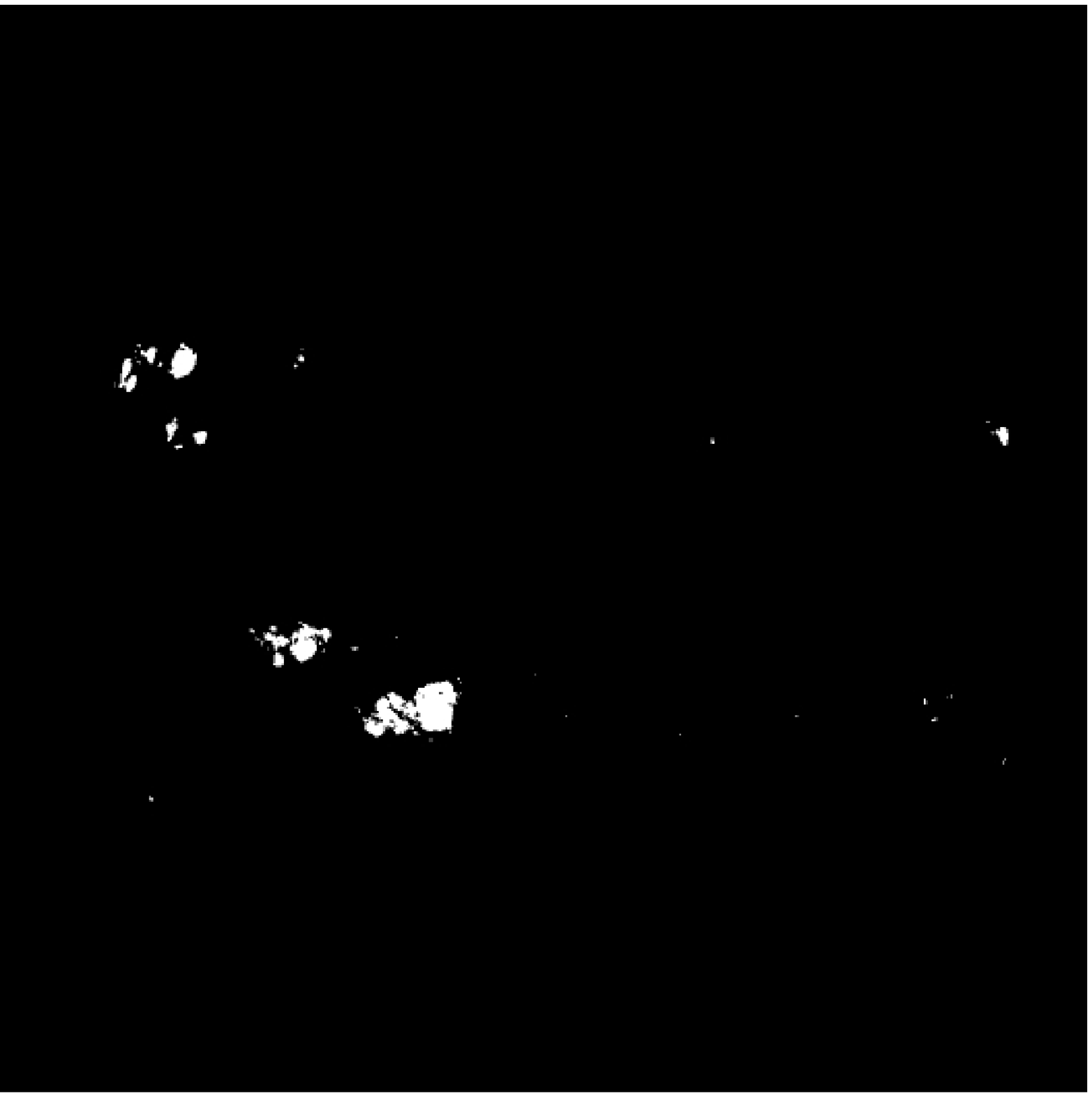}
			\caption{\label{fig:ss}
		Left panel: MDI intensitygram obtained May 17, 2000 at 11:11 UT. Middle panel: Sunspot segmentation map, which was achieved by applying Equation\,\ref{E:stats}. This is the results of the first stage of the initial sunspot segmentation. Right panel: Sunspot segmentation map, which was achieved by applying Equation\,\ref{E:pore}. This is the results of the second stage of the initial sunspot segmentation. The segmentation map from the second stage shows further detection of sunspots.}
\end{figure*}

For comparison purposes, we found that the corresponding value of this threshold to the original intensitygrams in FITS format is equal to 0.91. An example of the initial sunspot detection of an intensitygram using Equation\,\ref{E:stats} is shown in Figure\,\ref{fig:ss}.

\subsubsection{Pore Intensities}

\begin{figure*}[t!]
	\centering
	\includegraphics[width=0.80\textwidth]{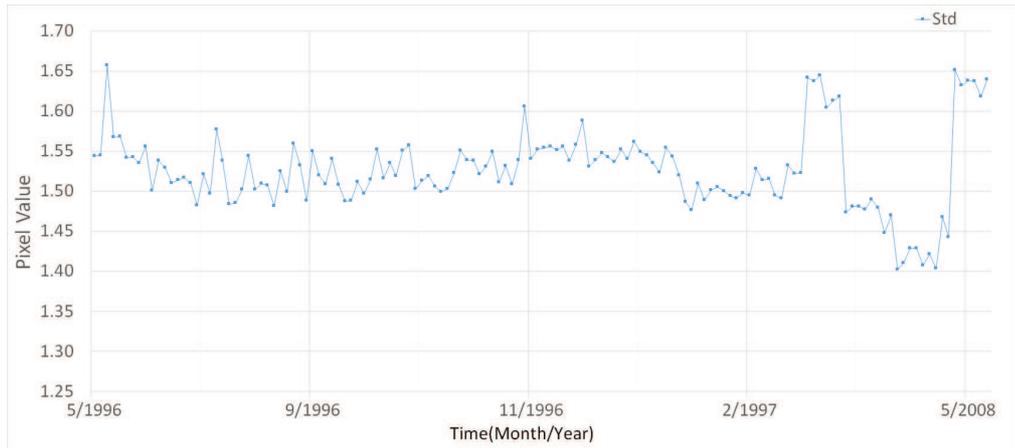}
	\caption{\label{fig:5}
		The standard deviation, $\sigma_{\rm I,SMin}$, of 140 intensitygrams collected from the solar minimum periods.}
\end{figure*} 

We carried out further analysis in order to enhance the detection of sunspots achieved from the first stage, as discussed in Section\,\ref{Sec:SSInt}, to detect smaller parts of sunspots and pores in each of the detected sunspot groups, by calculating $Th_{\rm Pore}$ using Equation\,\ref{E:pore}. 140 samples with the lowest activity during the solar minimum periods were collected and analysed at this stage in order to establish the values of the parameters in Equation\,\ref{E:pore}. The $\sigma_{\rm I}$ values of the collected data samples were analysed in order to determine an estimate of the standard deviation of the solar disk with the lowest activity ($\sigma_{\rm I,SMin}$). The collected data are plotted and shown in Figure\,\ref{fig:5}. When these samples were examined, care was taken to check that the solar disk was clear of sunspots and pores. The average and the median of $\sigma_{\rm I,SMin}$, $\left\langle \sigma_{\rm I,SMin} \right\rangle$ and $\left\langle \sigma_{\rm I,SMin} \right\rangle_{\rm med}$, were found to be equal to 1.5. We consider this value to reflect the standard deviation value of the quiet Sun at the solar minimum. Thus, $\sigma_{\rm I,SMin}$, $\left\langle \sigma_{\rm I,SMin} \right\rangle=1.5$, is adopted for the $Th_{\rm Pore}$ calculation.
We also carried out an empirical analysis on a number of sunspot detections in a 100$\times$100-pixel window, which is equivalent to a field-of-view of about 200$\times$200 arcsec. The aim was to find a threshold value, $Th_{\rm Pore}$, that can enhance the detection of sunspots, as defined in Equation\,\ref{E:pore}. The analysis was carried out by linearly increasing the value of $f_{\rm p}$ to find the most appropriate value for the detection of pores, which was found to be 7. Further discussion about enhancing the sunspot detection is given in Section\,\ref{Sec:MDISeg}. An example of this stage of sunspot detection in comparison with the results from the first stage, is shown in Figure\,\ref{fig:ss}. 

\begin{equation}
Th_{\rm Pore}= \left\langle I \right\rangle_{\rm region} - \left( f_{\rm p} \times \sigma_{\rm I,SMin} \right), \label{E:pore}
\end{equation}

where $f_{\rm p}=7$ and $\sigma_{\rm I,SMin}=1.5$.

\subsection{Magnetogram Analysis}\label{sec:maganal}
Having converted the magnetograms to unsigned 8-bit greyscale images, we record the mean $\left\langle B \right\rangle$, standard deviation $\sigma_{\rm B}$, minimum $B_{\rm min}$, and maximum $B_{\rm max}$, of the scaled on-disk magnetic field values for the whole dataset summarised in Table\,\ref{T:statsB} and the following subsections. 

\begin{table}[h!]
	\caption{Summary of the statistical analysis of the magnetograms. }
	\begin{center}
		{\footnotesize 
			\begin{tabular}{lcl}
				\hline
				\noalign{\smallskip}
				\noalign{\smallskip}
				Parameter & 8-bit Value & Description \\
				\noalign{\smallskip}
				\noalign{\smallskip}
				\hline
				\noalign{\smallskip}
				\noalign{\smallskip}
				
				$\left\langle \left\langle B \right\rangle \right\rangle$, $\left\langle \left\langle B \right\rangle \right\rangle_{\rm med}$  & 127 &  The mean and median value of  $\left\langle B \right\rangle$ during\\
				&  & the solar cycle. \\
				\noalign{\smallskip}
				\noalign{\smallskip}
				
				$\left\langle B \right\rangle_{\rm SMin} - \min (B_{\rm min, SMin})$ & 30 & The difference between $\left\langle B \right\rangle$ and $\min (B_{\rm min})$ \\
																					&  & during the minimum period of the solar cycle.\\
				\noalign{\smallskip}
				\noalign{\smallskip}																	
				$\left\langle B \right\rangle_{\rm SMin} - \max (B_{\rm max, SMin})$ & 30 & The difference between $\left\langle B \right\rangle $ and $\max (B_{\rm max})$ \\
																					&  & during the minimum period of the solar cycle.\\
																					
				\noalign{\smallskip}
				\noalign{\smallskip}																	
				$\sigma _{\rm B,SMin}$ & 2.4  & The standard deviation of the magnetic field strength  \\
									   &     & during the minimum period of the solar cycle.\\
				\noalign{\smallskip}
				\hline
			\end{tabular}
			\label{T:statsB}}
	\end{center}
\end{table}

\subsubsection{Magnetogram Statistics}
\begin{figure*}[t!]
	\centering
	\includegraphics[width=0.80\textwidth]{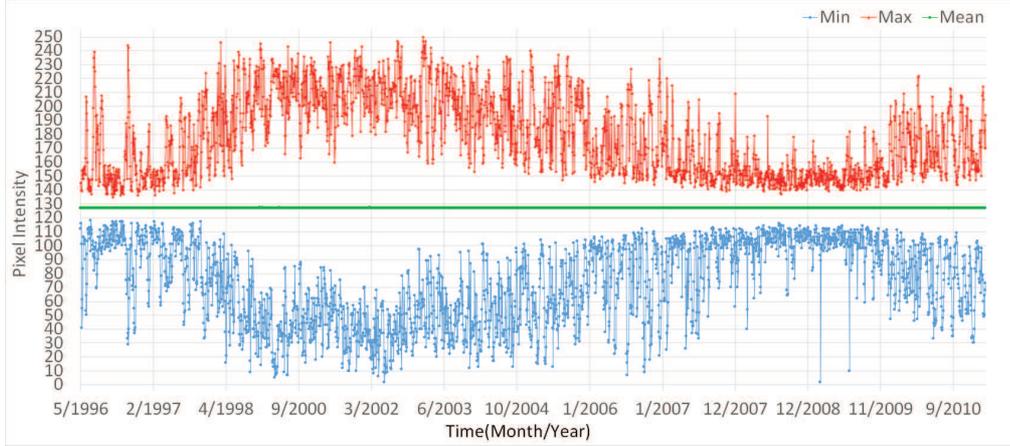}
	\caption{\label{fig:8}
	$\left\langle B \right\rangle$ (green), $B_{\rm min}$ (blue), and $B_{\rm max}$ (red) of the full disk for the full set of 8-bit magnetograms.}
\end{figure*}

We determined $\left\langle B \right\rangle$, $B_{\rm min}$, and $B_{\rm max}$ for the on-disk pixels of the full set of the analysed magnetograms. The collected data are plotted in Figure\,\ref{fig:8}. 

\begin{figure*}[t!]
	\centering
	\includegraphics[width=0.80\textwidth]{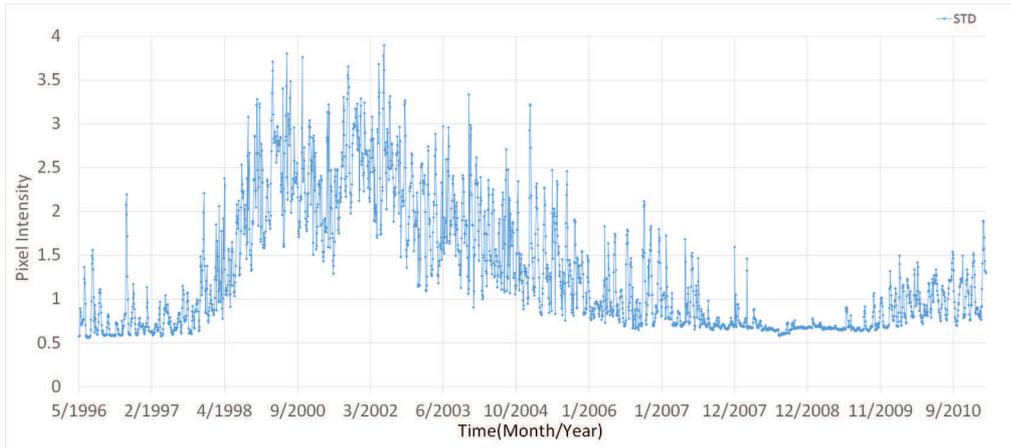}
	\caption{\label{fig:9}
		The standard deviation of the magnetic field strength, $\sigma _{\rm B}$, of the full disk for the full set of 8-bit magnetograms.}
\end{figure*}

We find the mean magnetic field strength $\left\langle B \right\rangle$ to be fairly constant over the full time series, and to be on average 127. The $B_{\rm min}$ and $B_{\rm max}$ values were also collected as they can give an indication of the intensities of the negative and positive polarity regions. As expected, $B_{\rm min}$ and $B_{\rm max}$ values vary in line with the solar cycle. As a result of the 8-bit rescaling of negative and positive magnetic field strength values onto one positive scale, the $B_{\rm min}$ values are higher during the solar minimum period and gradually decreases towards the solar maximum period, while $B_{\rm max}$ values are lower during the solar minimum period and gradually increase with solar activity. 

Furthermore, the standard deviation of the magnetic field strength, $\sigma _{\rm B}$, has also been determined for the full set of magnetograms and plotted in Figure\,\ref{fig:9}. As expected, we find $\sigma _{\rm B}$ to be lower during the solar minimum period and to increase with the presence of solar activity. In the following, we use $\sigma _{\rm B}$ as an indicator for finding magnetograms with the lowest solar activity in the data set.

\begin{figure*}[t!]
	\centering
	\includegraphics[width=0.80\textwidth]{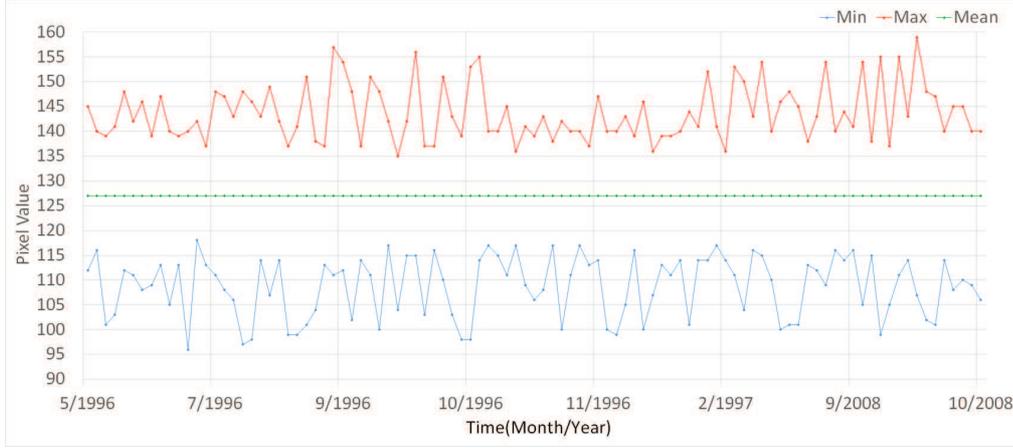}
	\caption{\label{fig:10}
		$\left\langle B \right\rangle$ (green), $B_{\rm min}$ (blue), and $B_{\rm max}$ (red) of 100 magnetograms, with the lowest $\sigma _{\rm B}$, collected from the solar minimum periods.}
\end{figure*}

\subsubsection{Magnetic Region Analysis}\label{sec:MRAnalysis}
During the solar minimum, the Sun has lower magnetic activity and flux when compared to the solar maximum period. In this context, we collected 100 samples of magnetograms from the solar minimum periods when solar activity was lowest and analysed the $\left\langle B \right\rangle$, $B_{\rm min}$, and $B_{\rm max}$ values, in order to get an initial estimation of the pixel values that represent magnetic regions. The collected data are shown plotted in Figure\,\ref{fig:10}. At this stage, we are unable to define the threshold values for detecting magnetic regions. However, we consider the average of the minimum, $\min(B_{\rm min})$, and the maximum, $\max(B_{\rm max})$, values of the collected samples to indicate magnetic values. We found the differences between these values and $\left\langle B \right\rangle$, $\left| \left\langle B \right\rangle - \min(B_{\rm min})\right|$, and $\left| \left\langle B \right\rangle - \max(B_{\rm max})\right|$, to be approximately 30 and -30. These values are adopted to enable the enhanced visualisation of the magnetic regions, which is achieved by applying contrast stretching, so that values between $\left\langle B \right\rangle$ - 30 and $\left\langle B \right\rangle$ + 30 will be transformed to values between 0 and 255 respectively. Equation\,\ref{E:CS} is applied to perform the contrast stretching, where in this case the interval of the magnetic field values to be stretched are $B^*_{\rm Min}=\left\langle B \right\rangle - 30$ and $B^*_{\rm Max}=\left\langle B \right\rangle + 30$, respectively. 

\begin{equation}
B_{\rm Output}= B_{\rm Input} - B^*_{\rm Min} \frac{255}{B^*_{\rm Max}-B^*_{\rm Min}}, \label{E:CS}
\end{equation}

\begin{figure*}[t!]
	\centering
	\includegraphics[width=0.6\textwidth]{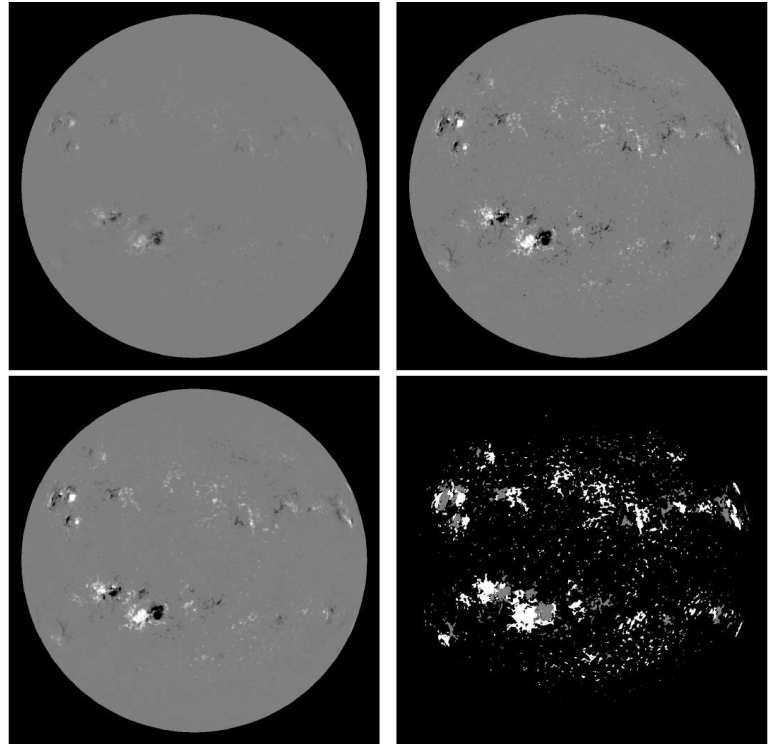}
	\caption{\label{fig:11}
		Top left panel: MDI magnetogram obtained May 17, 2000 at 01:39 UT. Values in the range of -3,500 to +3,500 Gauss are represented in grey-scale. Top right panel: the contrast stretched version of the magnetogram. Values with the equivalence of -823 to +823 Gauss are represented in grey-scale. Bottom left: median filtered version of the contrast stretched magnetogram. Bottom right: magnetic region segmentation map. The grey and white regions on the segmented map represent the negative and positive regions on the magnetogram, respectively. }
\end{figure*}

After performing the contrast stretching process, we carried out investigations to find the most appropriate approach for detecting magnetic regions within the solar disk, using the investigated samples. However, as reported in \cite{Ortiz2002} and \cite{Liu2012SoPh}, we find that the magnetograms have non-uniform noise, and the application of a constant threshold results in the detection of significant numbers of isolated pixels particularly in the lower right quadrant of solar disks, where the actual noise is slightly above the average used in the thresholding. This segmented noise has been eliminated by applying a 3$\times$3 median filter to the magnetogram, with little effect on the segmented signal. An example of an enhanced magnetogram is shown in Figure\,\ref{fig:11}.

After enhancing the investigated images, we carried out an investigation on the collected data in order to determine the threshold values for the magnetic regions detection ($Th_{\rm mr}$). We did so by investigating the most appropriate value for the variable $f_{\rm mr}$ in Equation\,\ref{E:5} that can be used to detect the magnetic regions. The value of $\sigma_{\rm B,SMin}$ in Equation\,\ref{E:5} is a predetermined value, fixed at the value of $\sigma_{\rm B}$ of the solar disk at the solar minimum, which is approximately equal to 2.4. This value was determined by calculating the  $\sigma_{\rm B}$ of the contrast stretched magnetogram samples which were collected from the solar minimum periods, when magnetic activities were the lowest. Thus  $\sigma_{\rm B,SMin}$ is set to 2.4 for the magnetic region detection and with this we found, by visual inspection, that using the value 2 for $f_{\rm mr}$ gave the most appropriate results. For purpose of comparison, we found the physical equivalent of $Th_{\rm mr}$ to be about 30 Gauss. This is equivalent to about three times the noise level of the MDI magnetograms, as found by other studies (e.g \cite{Liu2012SoPh}). Figure\,\ref{fig:11} shows an example of a magnetic region segmentation map, which was achieved by applying Equation\,\ref{E:5}: 
  
\begin{equation}
Th_{\rm mr}= \left\langle B \right\rangle \pm \left( f_{\rm mr} \times \sigma_{\rm B,SMin}\right) , \label{E:5}
\end{equation}
where $f_{\rm mr}$=2, and $\sigma_{\rm B,SMin}$=2.4 in the 8-bit scale of the magnetograms.

\section{MDI Segmentation}\label{Sec:MDISeg}
ASAP is the collective name for a set of algorithms used to process solar images. These cover algorithms for detecting sunspots and active regions \citep{ColakQahwaji2008}, solar-flare prediction \citep{ColakQahwaji2009} and sunspot detection and tracking  \citep{Verbeeck2014}. For the purposes of the work described here, new segmentation algorithms were developed and integrated into ASAP to enable the detection and identification of faculae and network thought to be important for irradiance reconstruction. 
In summary, we apply our segmentation algorithms as follows. First the initial feature detection processes, described in Section\,\ref{sec:initial}, are applied to determine sunspot and magnetic region candidates. Second, a Neural Network (NN) method is applied, described in Section\,\ref{sec:NN}, to group sunspots and identify active regions based on magnetic region and sunspot properties. This step determines active region and their boundaries, which is a crucial intermediate process needed to determine whether features are located either inside or outside active regions. Finally, in Section\,\ref{sec:faculae}, the maps generated are analysed to identify faculae and network, based on their contrasts and proximity to active regions. In Section\,\ref{sec:alg}, we summarise the process and provide flowcharts of the overall segmentation process. 

\subsection{Initial Feature Detection}\label{sec:initial}
We process MDI intensitygrams and magnetograms for the initial detection of sunspots and magnetic regions. This is carried out in a similar manner to that described in \cite{ColakQahwaji2008} using automatically calculated threshold values, based on image type and the solar features that need to be detected. In this work, we adopted a number of predefined parameters, which are derived from analysing the MDI solar cycle data, as described in Section\,\ref{sec:dataanalysis}, and use them systematically in the segmentation approach, in order to achieve consistent segmentations. The initial detection of sunspots and magnetic regions are explained further below.

\subsubsection{Initial Sunspot Detection}
\begin{figure*}[t!]
	\centering
	\includegraphics[width=0.6\textwidth]{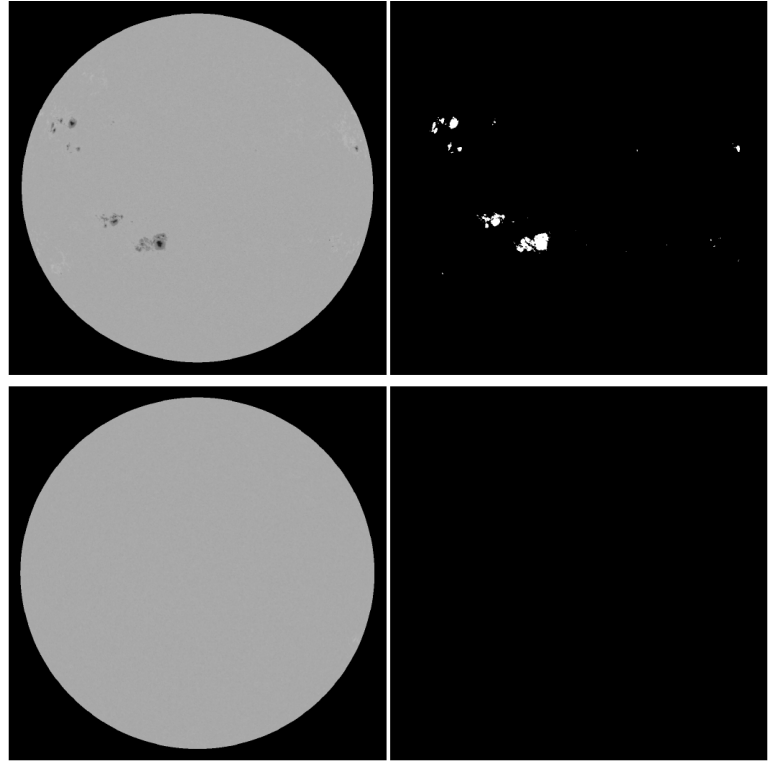}
	\caption{\label{fig:12}
		Left panel: intensitygrams from the solar maximum (top left) and solar minimum (bottom left). Right panel: initial sunspot segmentation maps from solar maximum (top right) and solar minimum (bottom right).}
\end{figure*}

The initial detection of sunspots includes two stages. First, in the MDI intensitygram, we detect pixels with values below ($\left\langle I \right\rangle - f_{\rm ss}$), which detects the majority of sunspot pixels, but occasionally misses smaller parts within the sunspots. It has been found that some of these missing small parts can be detected by the second stage, where a window of size 100$\times$100 pixels are drawn around each group of the detected pixels, and the mean is calculated for each window, which we denote here as ($\left\langle I \right\rangle_{\rm Region}$). Any intensitygram pixel within the window that has value lower than ($\left\langle I \right\rangle_{\rm Region} - (f_{\rm p} \times \sigma_{\rm I,SMin})$) is detected. Examples of sunspots segmentation maps are shown in Figure\,\ref{fig:12}.

\subsubsection{Initial Magnetic Region Detection}\label{sec:initialMR}
\begin{figure*}[t!]
	\centering
	\includegraphics[width=0.6\textwidth]{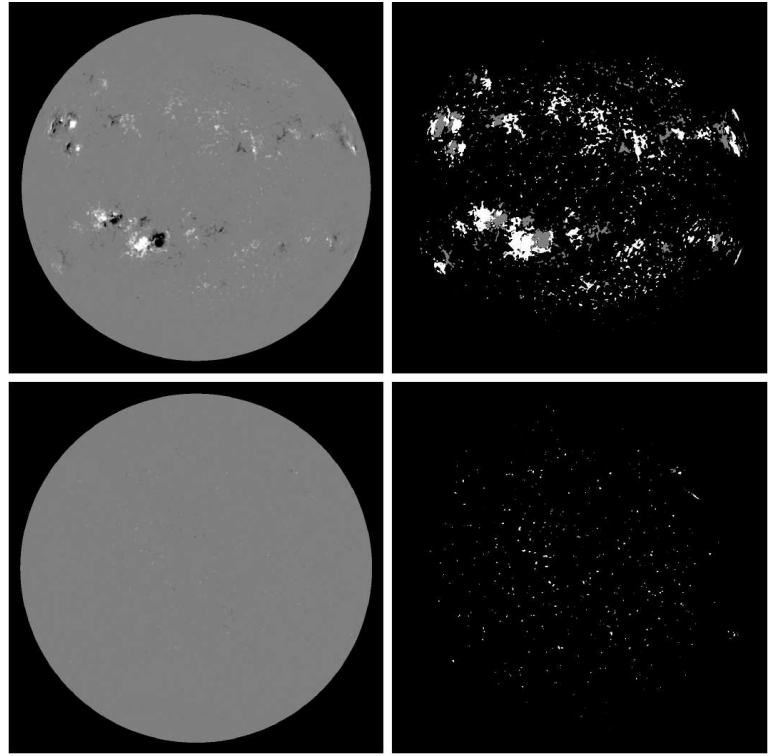}
	\caption{\label{fig:13}
		Left panel: magnetograms from the solar maximum (top left) and solar minimum (bottom left). Right panel: magnetic region segmentation maps from solar maximum (top right) and solar minimum (bottom right). The negative and positive regions on the magnetogram images are represented as grey and white regions respectively on the segmentation map, for clarity purpose.}
\end{figure*}

For the purpose of this paper, magnetic regions are defined as regions of the magnetograms that have magnetic field significantly higher than their surroundings. Prior to the detection of magnetic regions, we apply contrast stretching in order to enhance the visualisation of the magnetic regions, followed by 3$\times$3 window median filtering, to reduce the non-uniformed noise within the magnetogram, as described in Sec. \ref{sec:MRAnalysis}. Magnetic regions with positive and negative polarity are identified and segmented separately. For the segmentation threshold calculation, pixels that have values below ($\left\langle B \right\rangle - \left( f_{\rm m} \times \sigma_{\rm B,SMin} \right) $) represent negative polarity regions, while pixels that have values above ($\left\langle B \right\rangle + \left( f_{\rm m} \times \sigma_{\rm B,SMin} \right) $) represent positive polarity regions. In both cases, the $\sigma_{\rm B,SMin}$ value is set to 2.4, which is the average of $\sigma_{\rm B}$ in the solar minimum period when no activity was present, as discussed in Section\,\ref{sec:maganal}. Two examples of magnetic region detection from magnetograms are shown in Figure\,\ref{fig:13}.

\subsection{Final Feature Detection}
After detecting sunspot candidates and magnetic regions, the results are processed to cluster sunspots into groups, and also to identify groups of magnetic regions that overlap with sunspots, which we define as active regions. 

In this section, we summarise the determination of active region candidates using image processing, which is described in Section\,\ref{sec:arcan}, and the validation and final identification of active region and sunspot groups using neural network machine learning, which is described in Section\,\ref{sec:NN}. Full details of the sunspot and active region grouping can be found in \cite{ColakQahwaji2008}. We provide a summary of this process below. 
\subsubsection{Active Region Candidates}\label{sec:arcan}
In order to identify sunspot and active region groups, the magnetic region and sunspot candidate maps are combined as follows. First a pixel marked as a sunspot candidate ($P_{\rm spotcan}$) on the sunspot candidate map is found. If the magnetic region map has a magnetic pixel ($P_{\rm mag}$), at the same location as $P_{\rm spotcan}$, a new map is created for the active region group candidate ($P_{\rm actcan}$) and marked with the pixel value of the $P_{\rm mag}$. Then, all adjacent pixels to $P_{\rm actcan}$ that are marked as magnetic in the $P_{\rm mag}$, are marked as $P_{\rm actcan}$. 

\subsubsection{Neural Network for Active Region and Sunspot Group Identification}\label{sec:NN}
For the active region detection, we employ the machine learning approach that was introduced in \cite{ColakQahwaji2008}. A neural network (NN) was trained to enable the detection and grouping of active regions, using sunspot and active region candidates properties. A NN is a machine learning algorithm that is commonly used for tasks such as data classification, clustering and prediction. The adopted NN has been trained on nearly 100 examples of sunspots and active regions properties. For the NN application, the properties of overlapping sunspot and active region candidates are extracted and fed to the NN as an input. These properties include geometrical and location information of the active regions candidates. The NN then provide an output decision, which represents whether the active region candidate is considered to be an active region or not. The identified active regions are then used for the final determination of sunspot groups and their boundaries. 

\begin{figure*}[t!]
	\centering
	\includegraphics[width=0.3\textwidth]{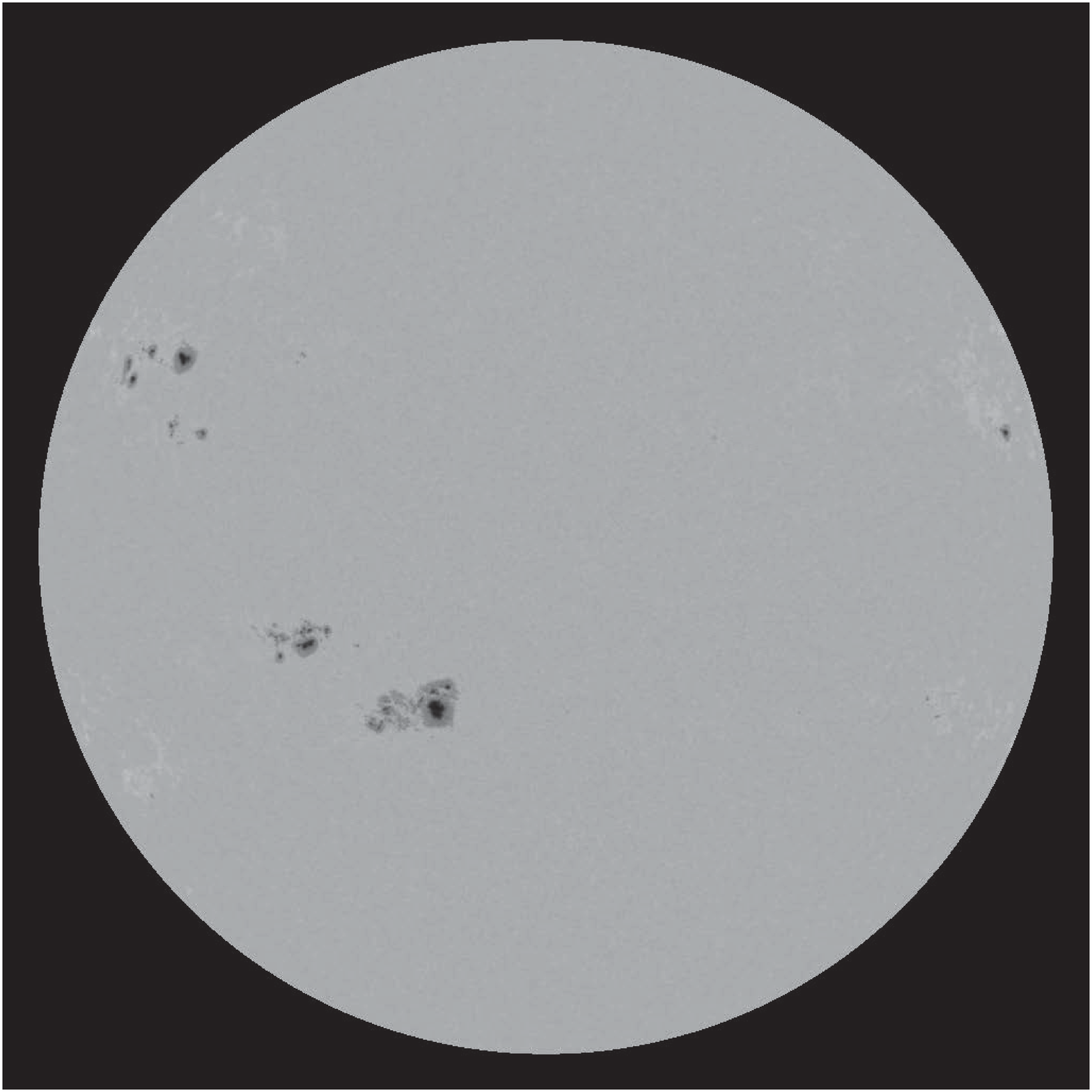}
	\includegraphics[width=0.3\textwidth]{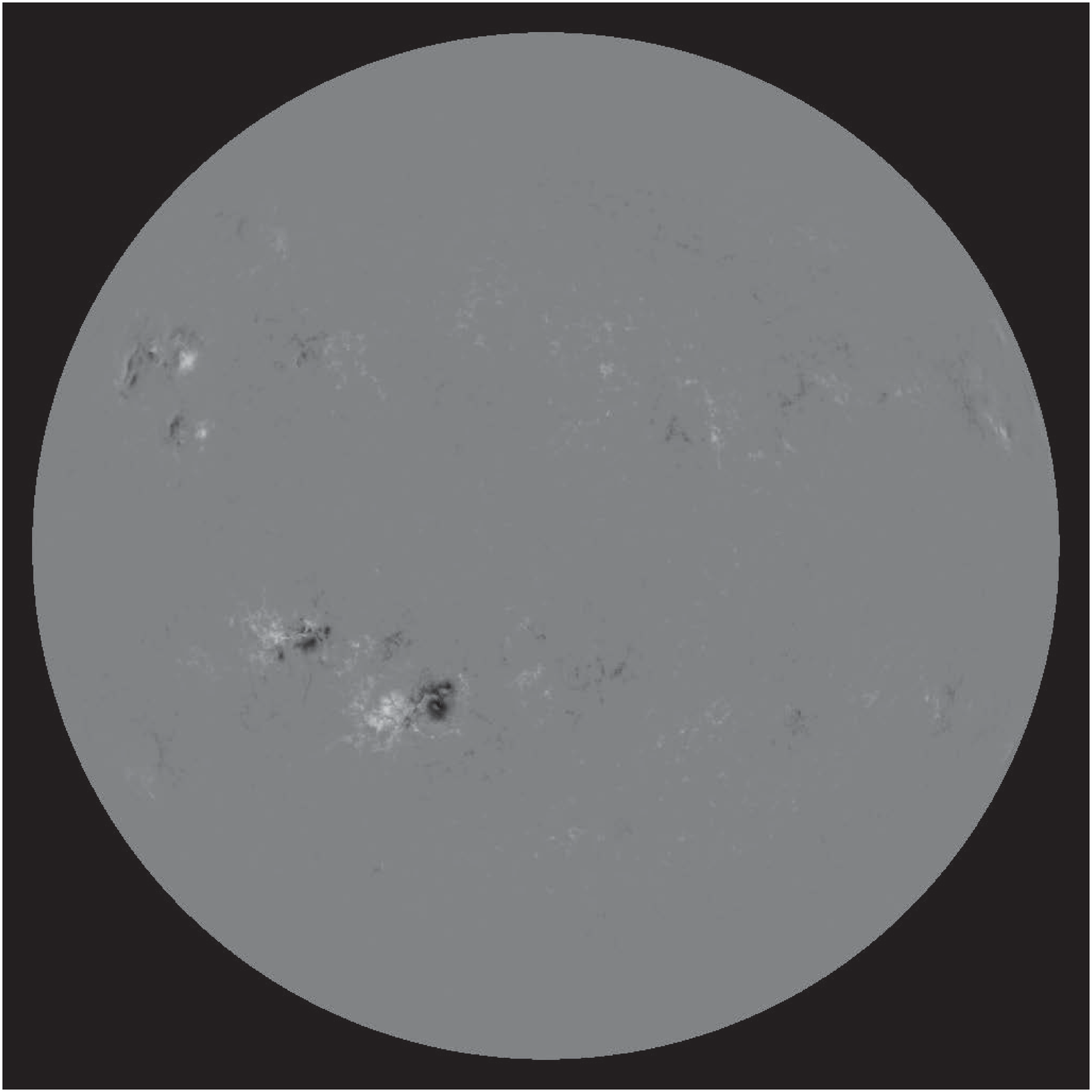}
	
	\includegraphics[width=0.3\textwidth]{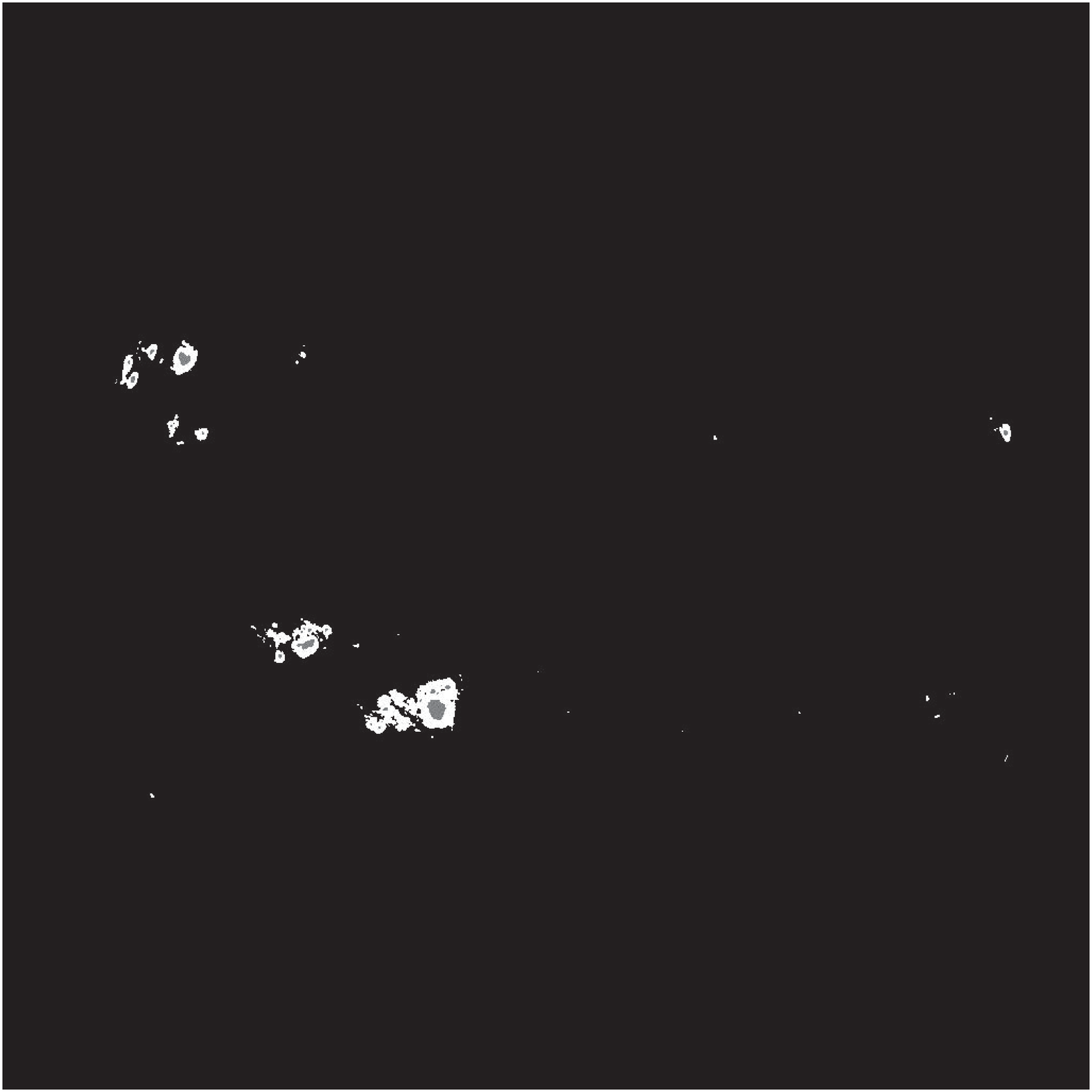}
	\includegraphics[width=0.3\textwidth]{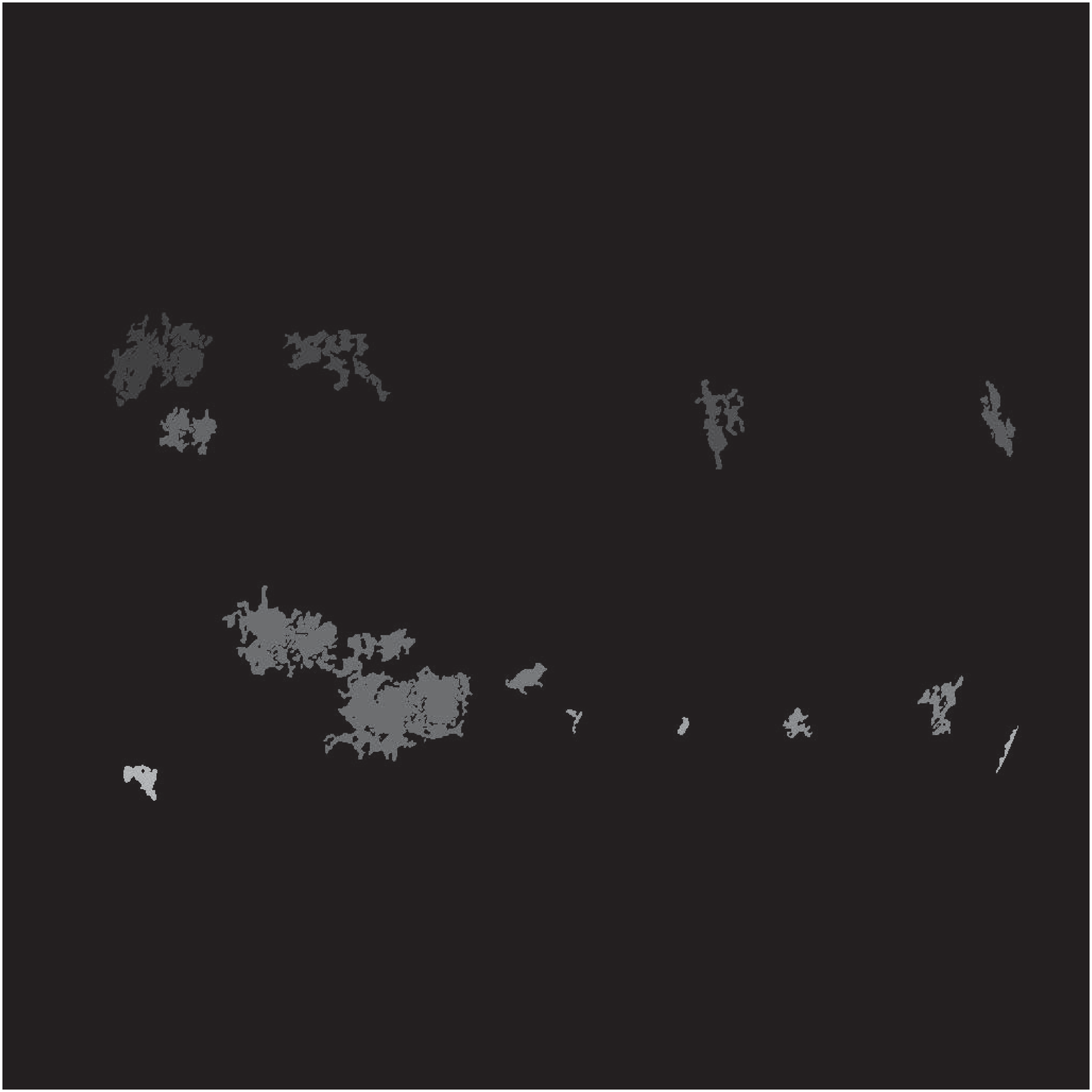}
	\caption{\label{fig:15}
		Top left and right, the input intensitygram and magnetogram. Bottom left: Map of sunspot umbra and penumbra. Bottom right: active region groups. Each group is given a different shade of grey.}
\end{figure*}

The umbra and penumbra parts of a sunspot are then detected, in a similar manner to the approach introduced in \cite{ColakQahwaji2008}. This approach classifies the sunspot pixels that have values below ($ \left\langle I_{\rm SS} \right\rangle  - \sigma_{\rm SS}$) as umbra, and the remaining sunspot pixels as penumbra. Here $ \left\langle I_{\rm SS} \right\rangle$ and $\sigma_{\rm SS}$ are the mean and the standard deviation, respectively, of the intensitygram pixels that are detected as sunspots. The final output from the processes discussed in this subsection are the sunspot groups, including umbra and penumbra, and active region groups, examples of which are shown in Figure\,\ref{fig:15}. 

\subsubsection{Final Detection of Faculae and Network}\label{sec:faculae}
\begin{figure*}[t!]
	\centering
	\includegraphics[width=0.3\textwidth]{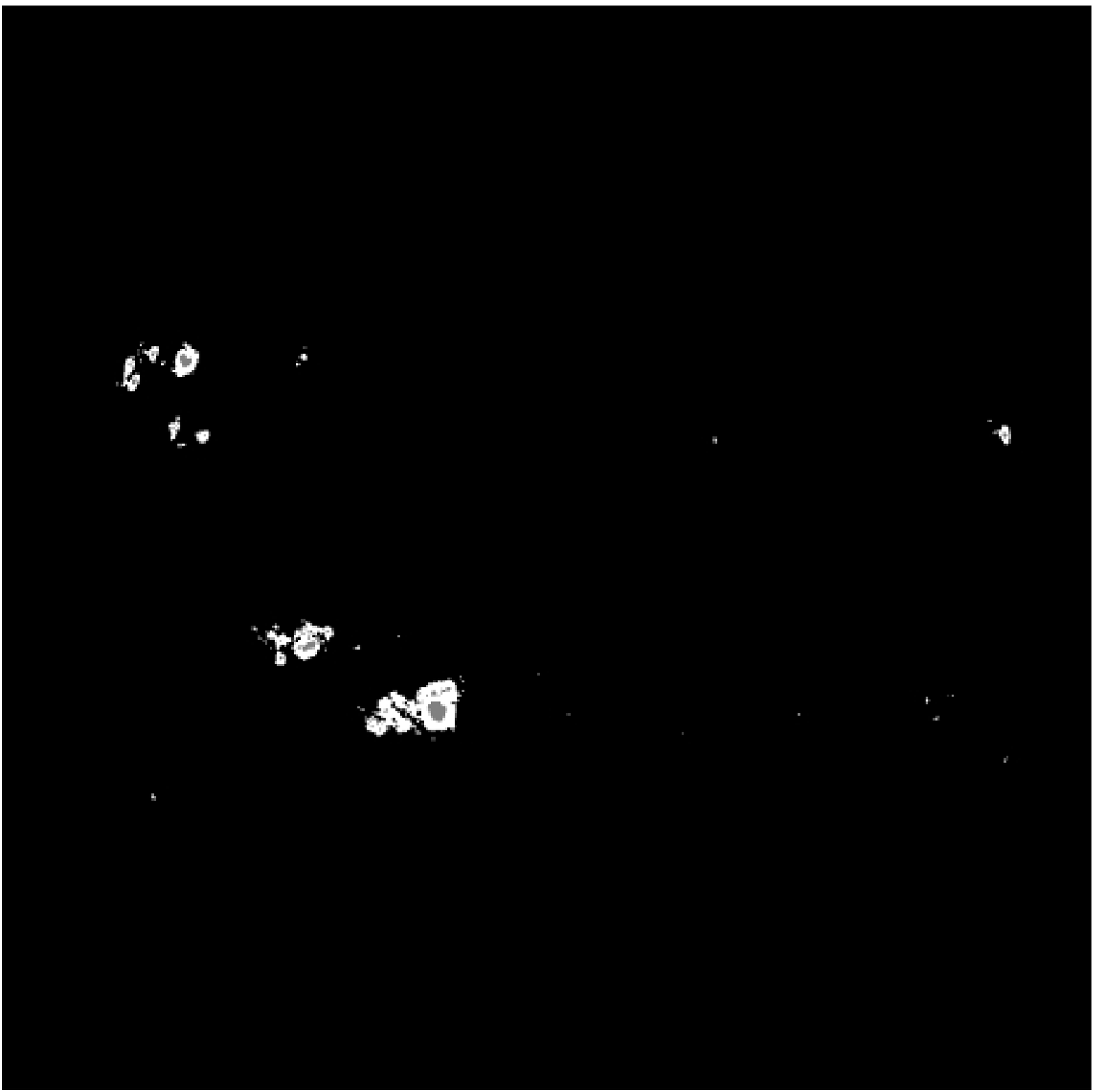}
	\includegraphics[width=0.3\textwidth]{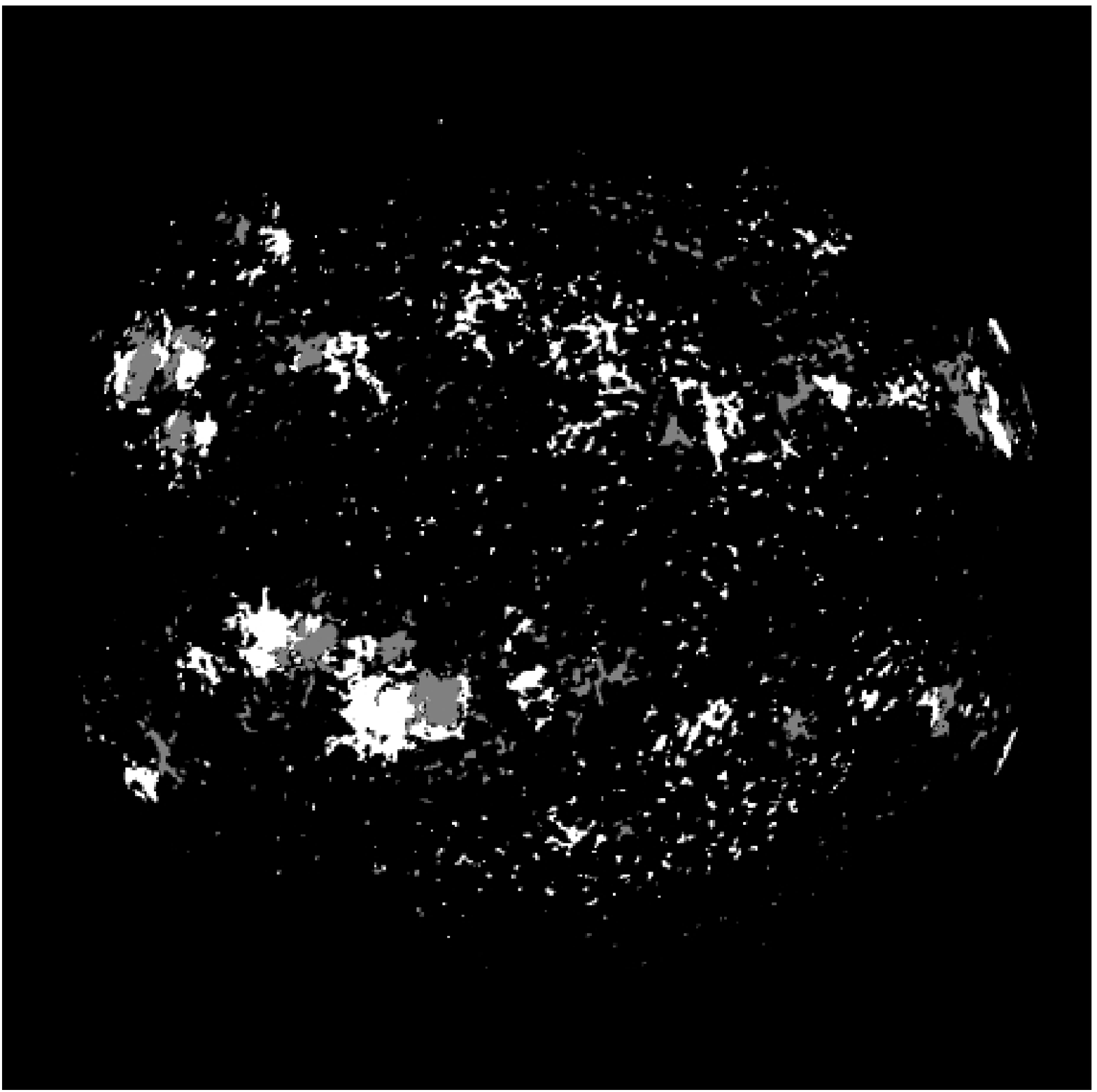}
	\includegraphics[width=0.3\textwidth]{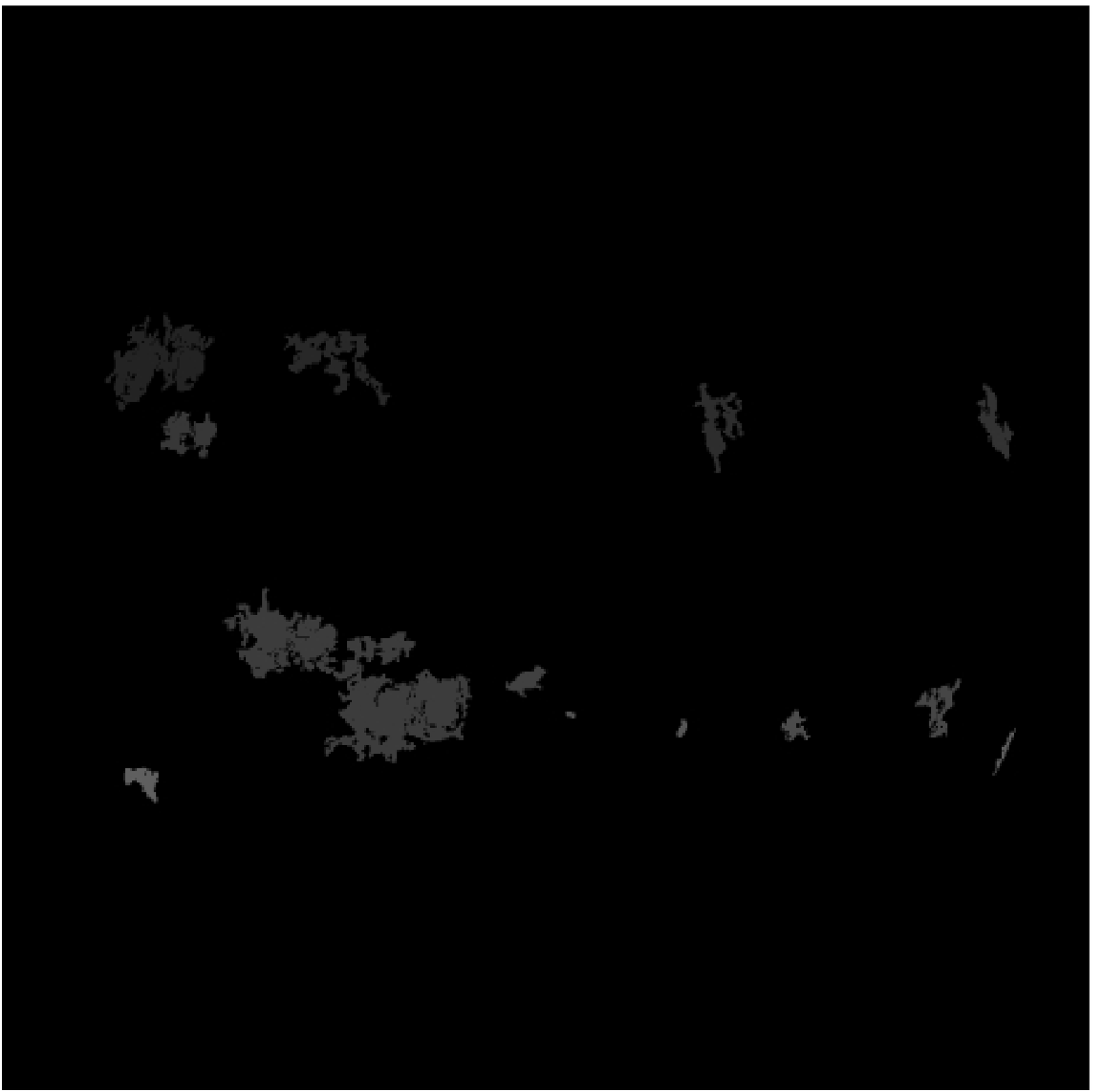}
							\caption{\label{fig:16}
		Example of the segmentation maps for sunspots (left), magnetic regions (centre), and active regions (right). These maps are further employed for the identification of faculae and network.}
\end{figure*}

 In this work, faculae are considered to be magnetic pixels that lie within the boundary of an active region, while network is defined as magnetic pixels located outside active region boundaries. For the distinction between faculae and network, ASAP’s active region detection is used as an intermediate step. The starting point to distinguish between these features is to employ image processing to the corresponding sunspot, magnetic region, and active region maps. An example set of these maps is shown in Figure\,\ref{fig:16}. 

\begin{figure*}[t!]
	\centering
	\includegraphics[width=0.3\textwidth]{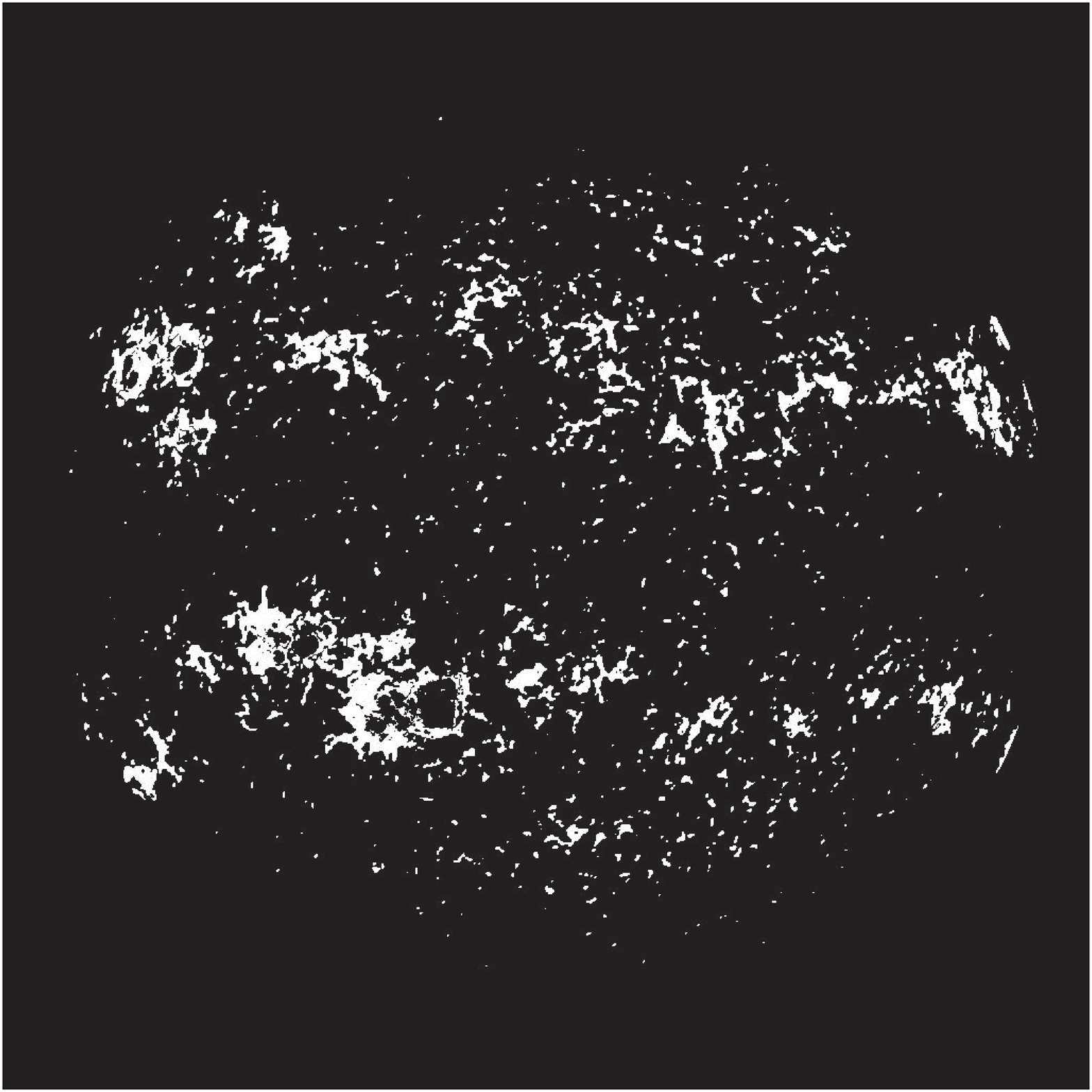}
	\includegraphics[width=0.3\textwidth]{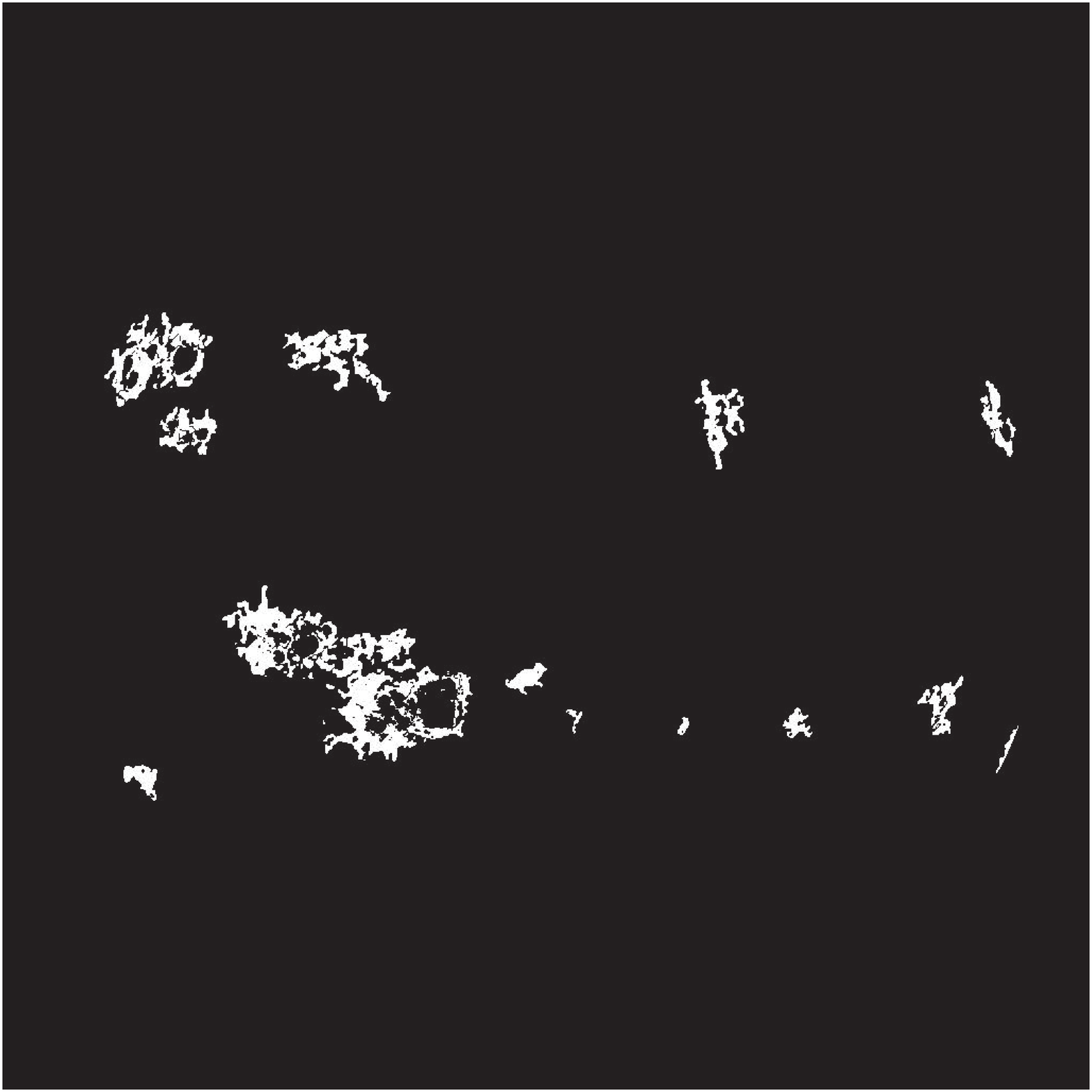}
	\includegraphics[width=0.3\textwidth]{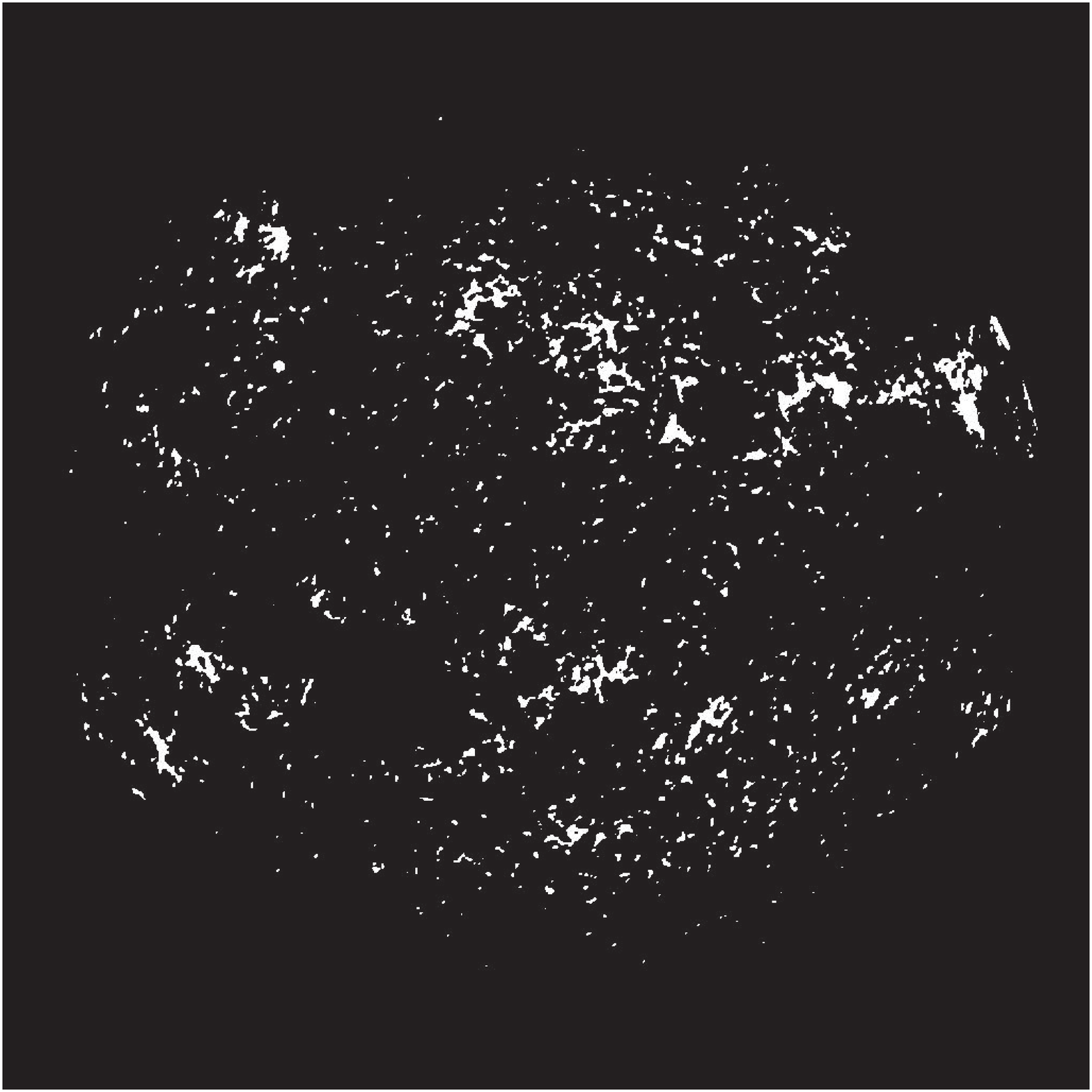}	
	\includegraphics[width=0.3\textwidth]{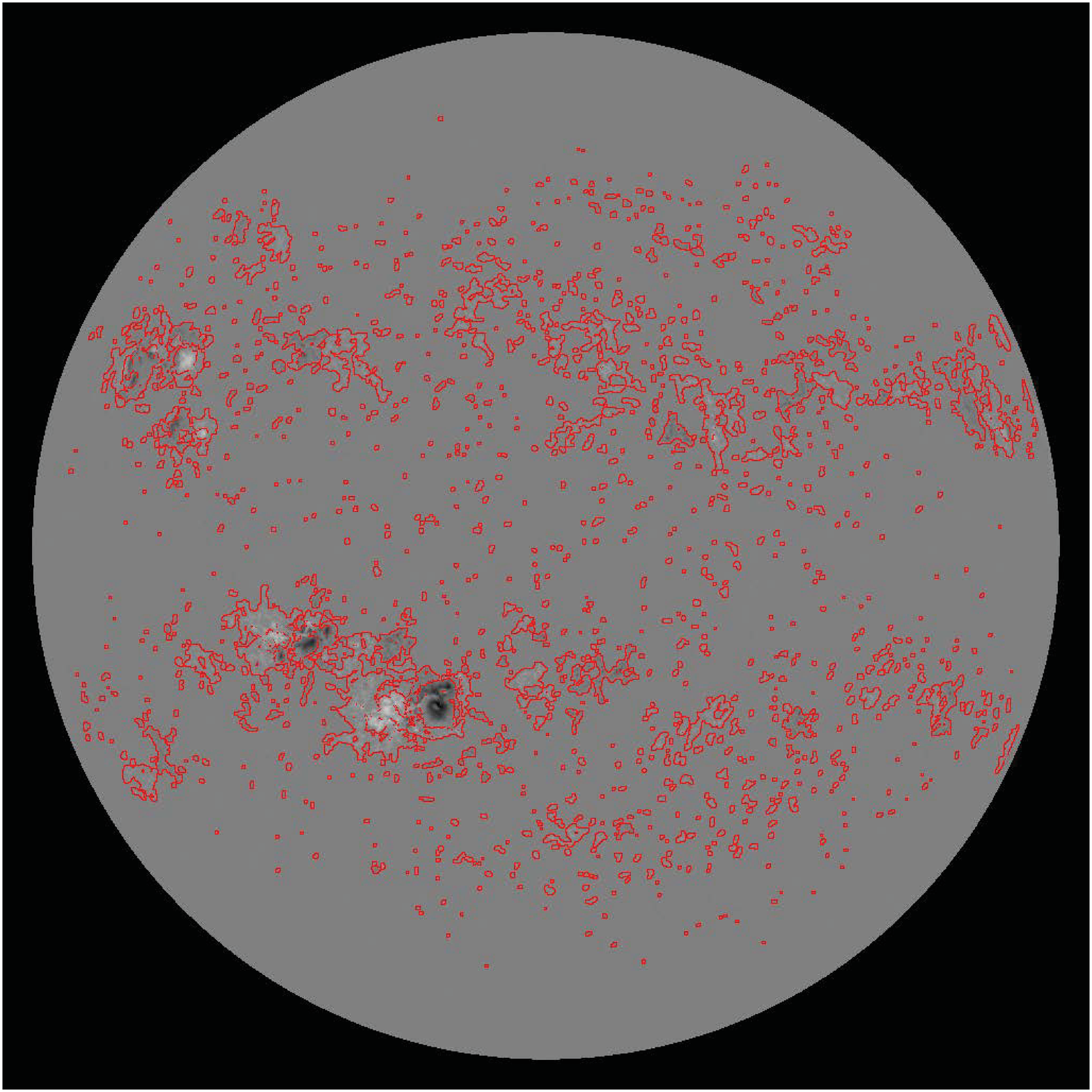}
	\includegraphics[width=0.3\textwidth]{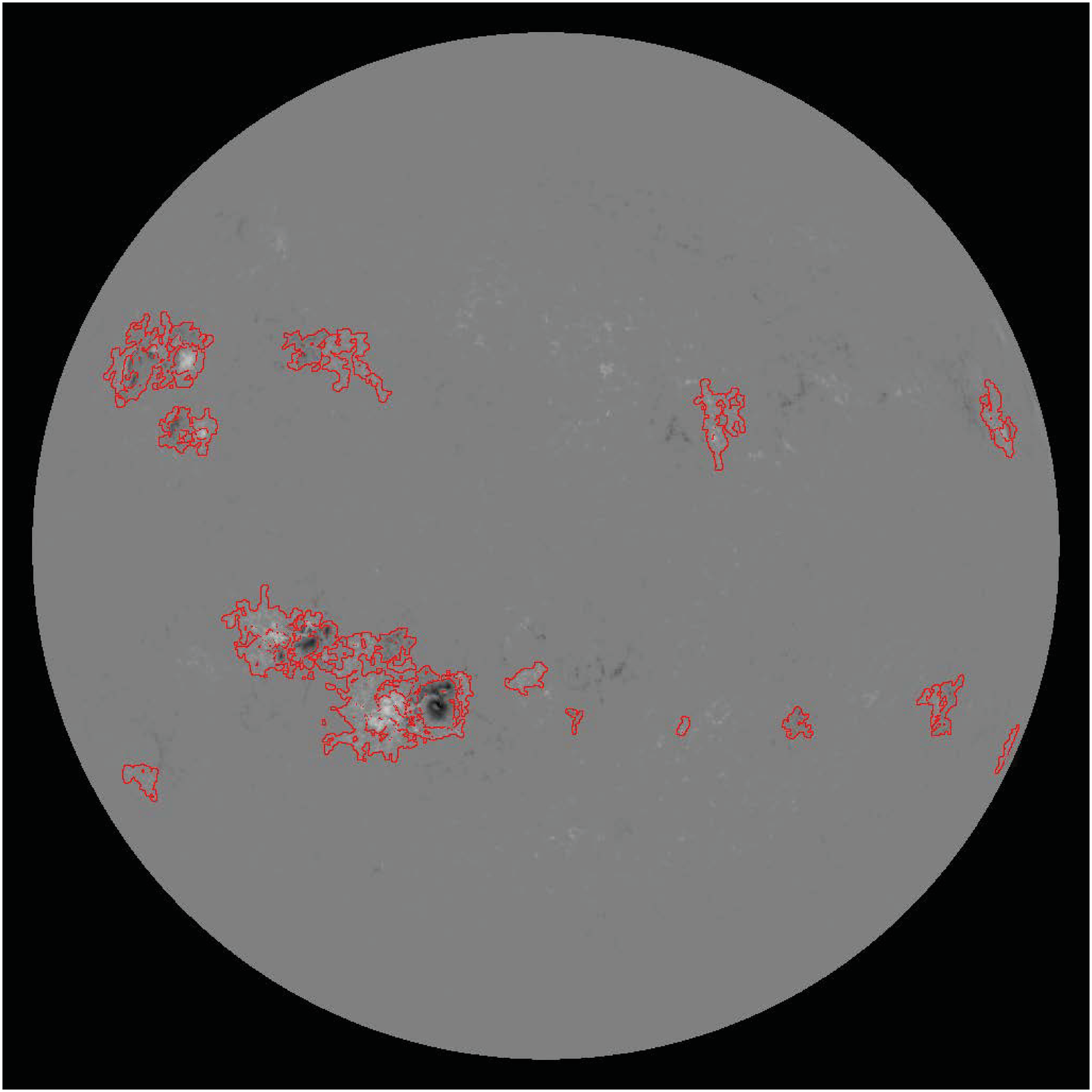}
	\includegraphics[width=0.3\textwidth]{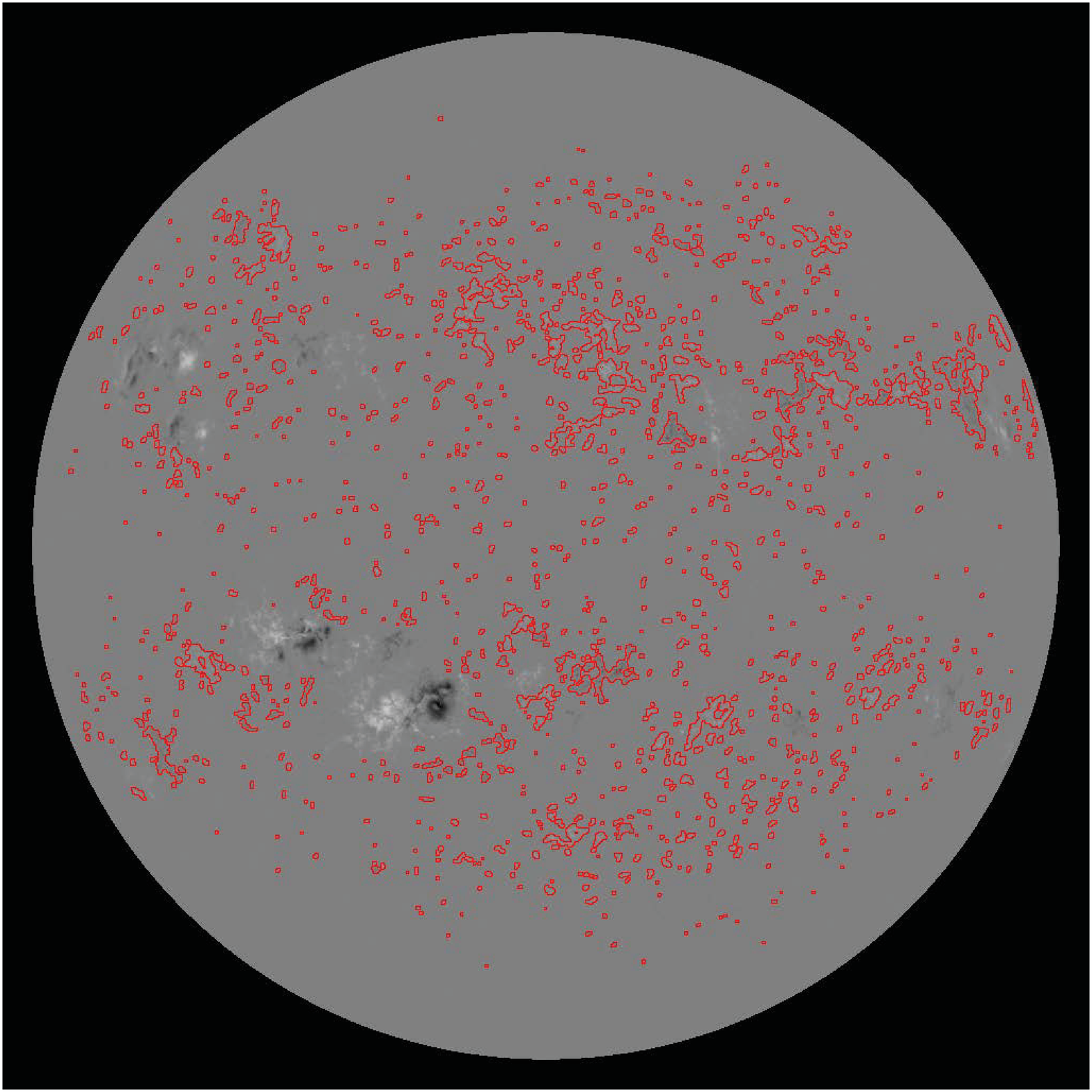}
	\caption{\label{fig:17}
		Segmentation maps are shown in the top row, from left to right, for faculae and network, faculae, and network with their corresponding boundaries shown in red in the bottom row for May 17, 2000 at 01:39 UT.}
\end{figure*}

\begin{figure*}[t!]
	\centering
	\includegraphics[width=0.3\textwidth]{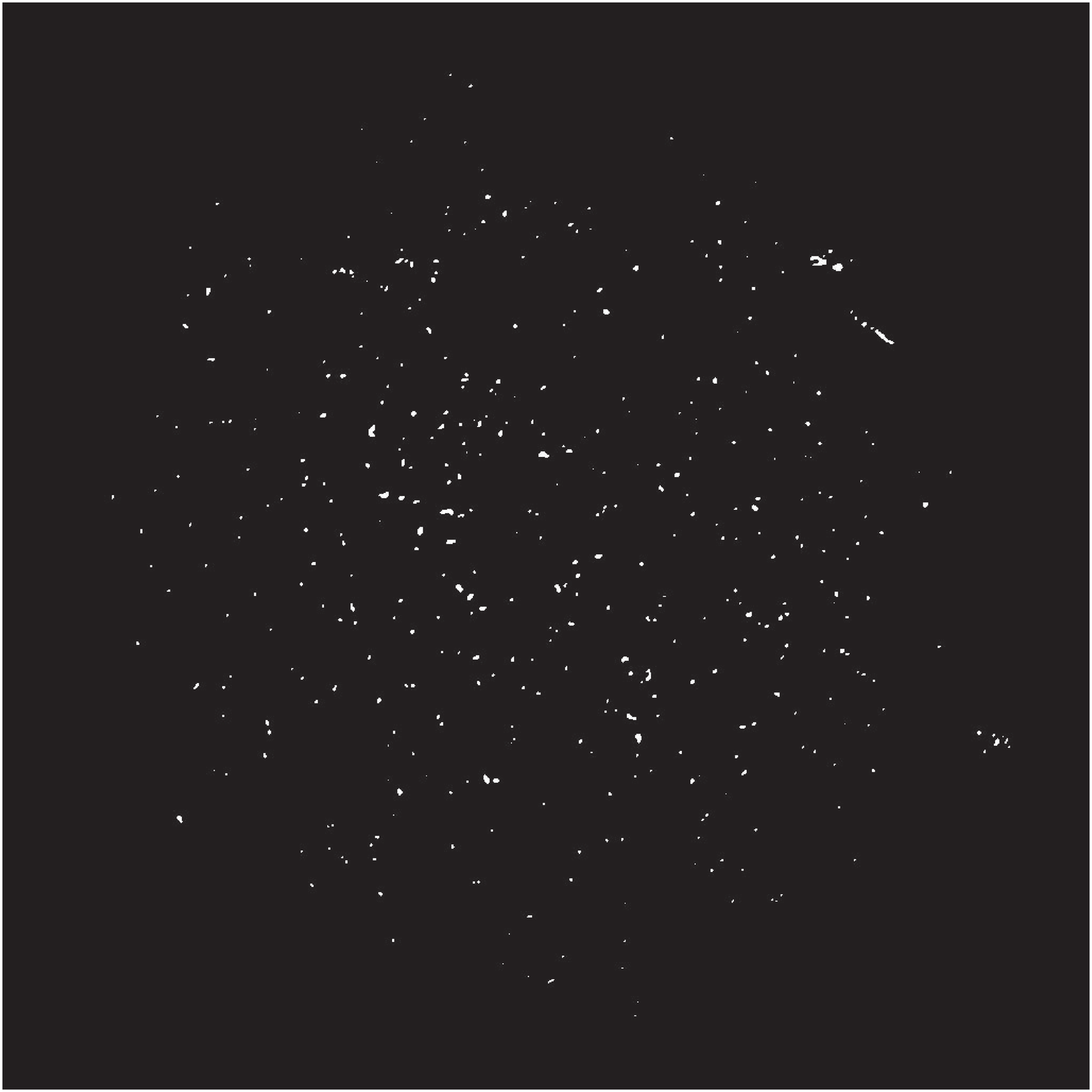}
	\includegraphics[width=0.3\textwidth]{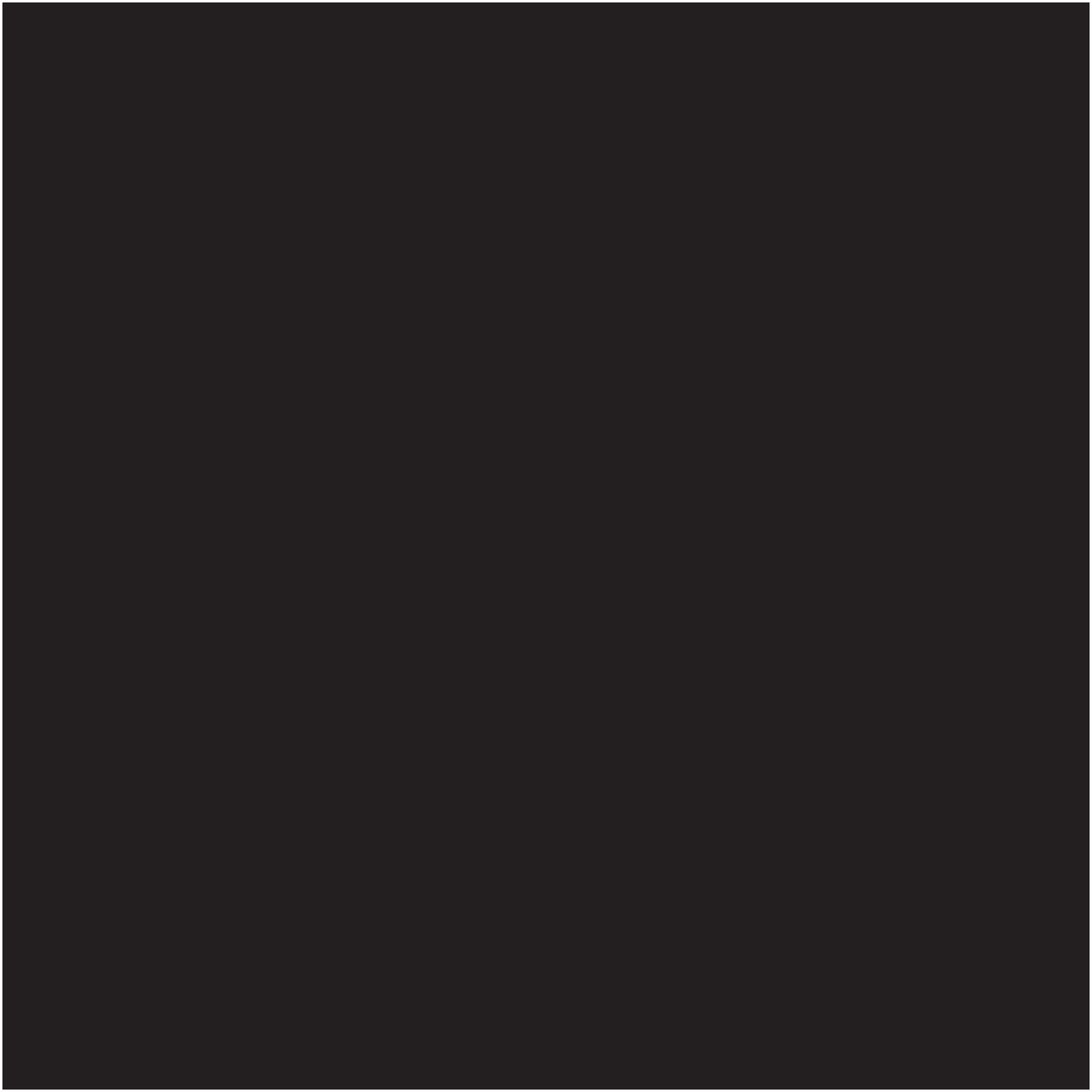}
	\includegraphics[width=0.3\textwidth]{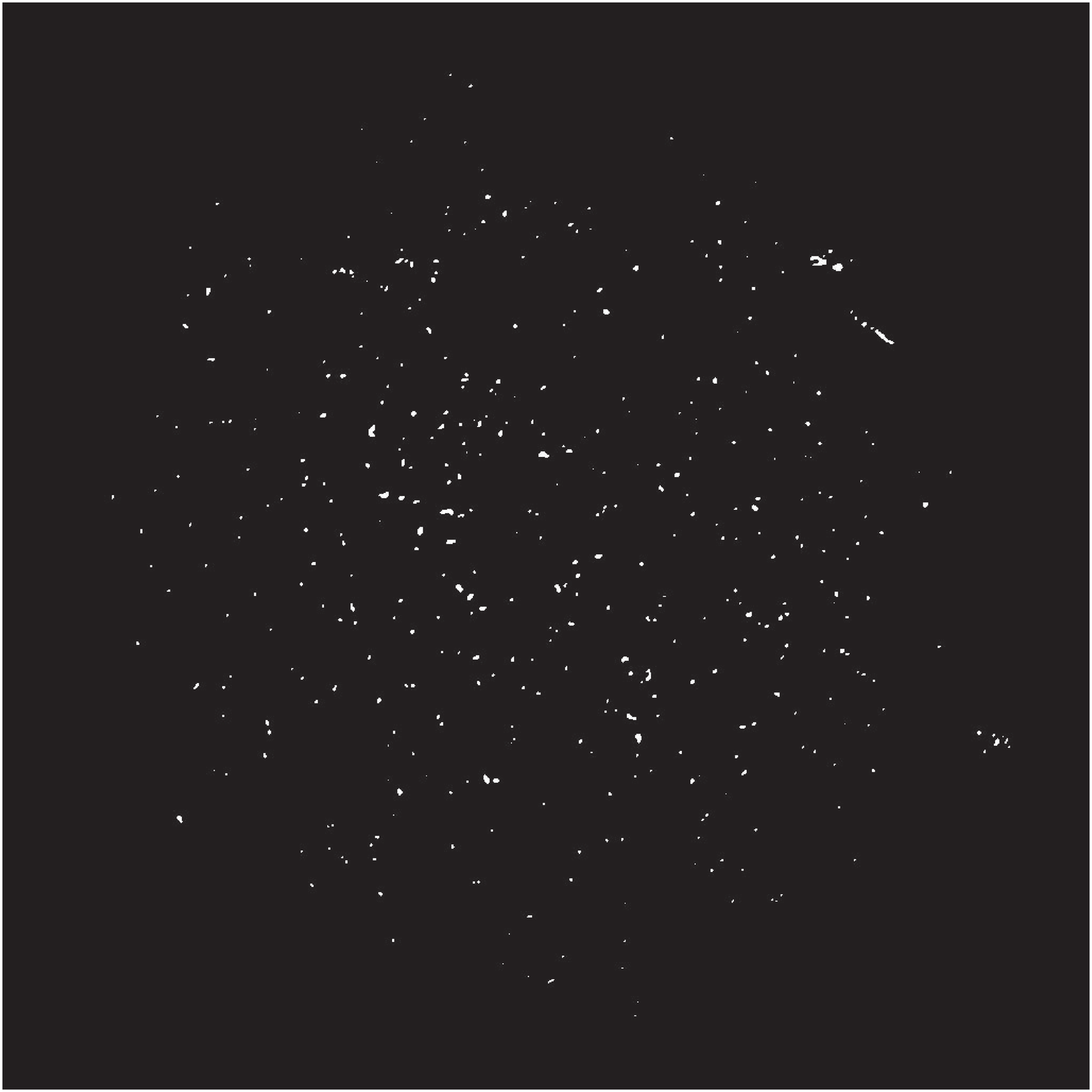}	
	\includegraphics[width=0.3\textwidth]{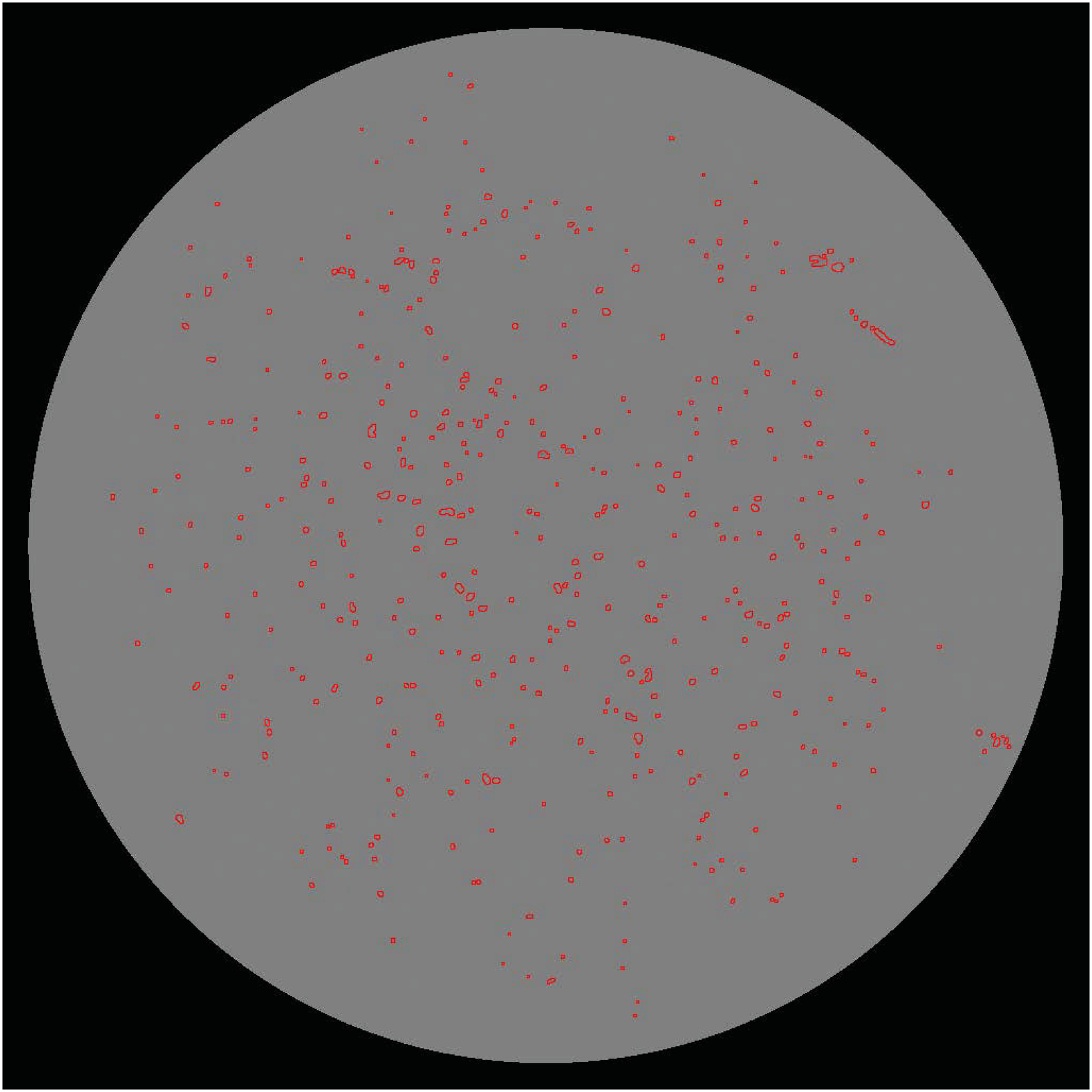}
	\includegraphics[width=0.3\textwidth]{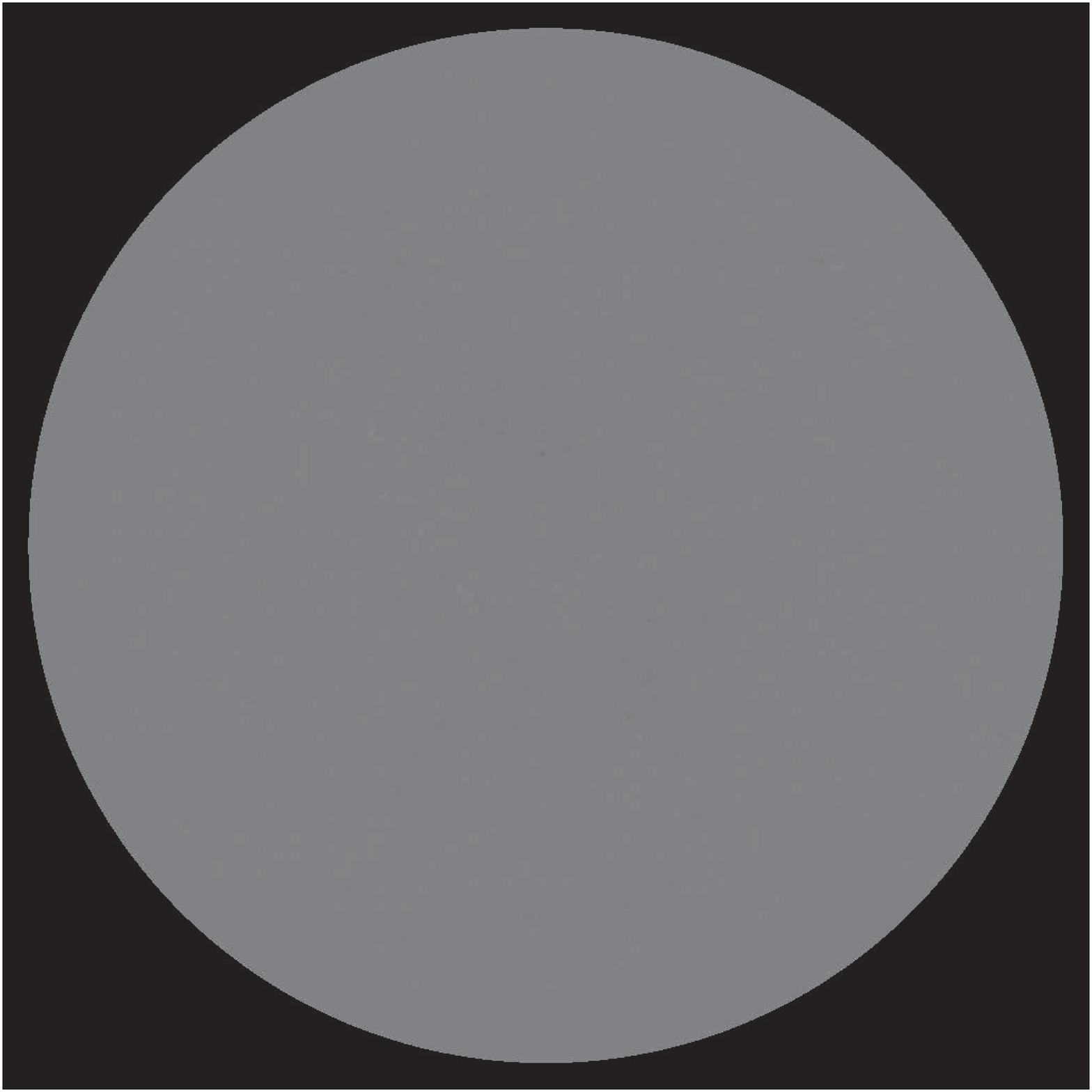}
	\includegraphics[width=0.3\textwidth]{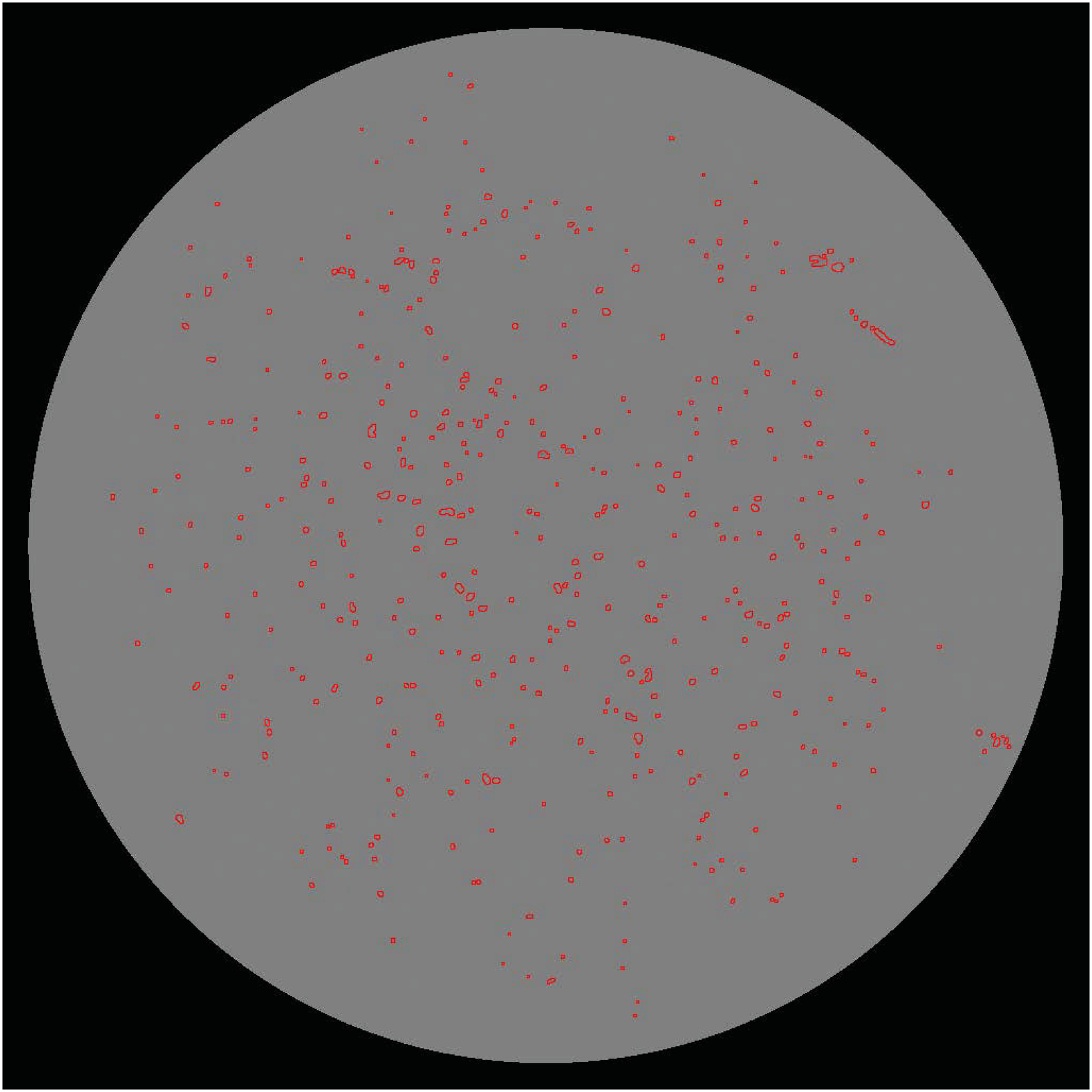}
	\caption{\label{fig:18}
		Segmentation maps are shown in the top row, from left to right, for faculae and network, faculae, and network with their corresponding boundaries shown in red in the bottom row for April 22, 1997, 00:04 UT. For this input, no faculae is detected, therefore no indication for faculae is given in the central segmentations.}
\end{figure*}

We identify the faculae and network map by subtracting the sunspot map from the magnetic region map. Then, the faculae and network pixels are sub-classified to faculae if they are located within active region, or to network if they are located outside an active region. Examples of detection results are shown in Figures\,\ref{fig:17} and \ref{fig:18}.

\cite{Ortiz2002} study the position-dependent contrast of magnetic elements in specific magnetic field strengths intervals from the threshold value up to 600\,G (see their Figure 3).  From this study it can be concluded that faculae at disk center with field strengths between 300 and 500\,G typically have a contrast $I_{\rm f}/I_{\rm qs}\approx$0.8 (or -0.2 in their representation), and a slightly lower contrast $I_{\rm f}/I_{\rm qs}\approx$0.75 for the interval of 500-600\,G. These contrast values are below the threshold of 0.91 used for sunspot identification in the present work. In order to ensure that faculae are not erroneously identified as sunspots, we included a criterion that sunspot pixels that have a magnetic field strength lower than 800\,G are always identified as facular pixels. Using this value we follow the approach by \cite{Ball2012}.

\subsection{Summary of Detection Algorithm}\label{sec:alg}
\begin{figure*}
	\centering
	\includegraphics[width=0.5\textwidth]{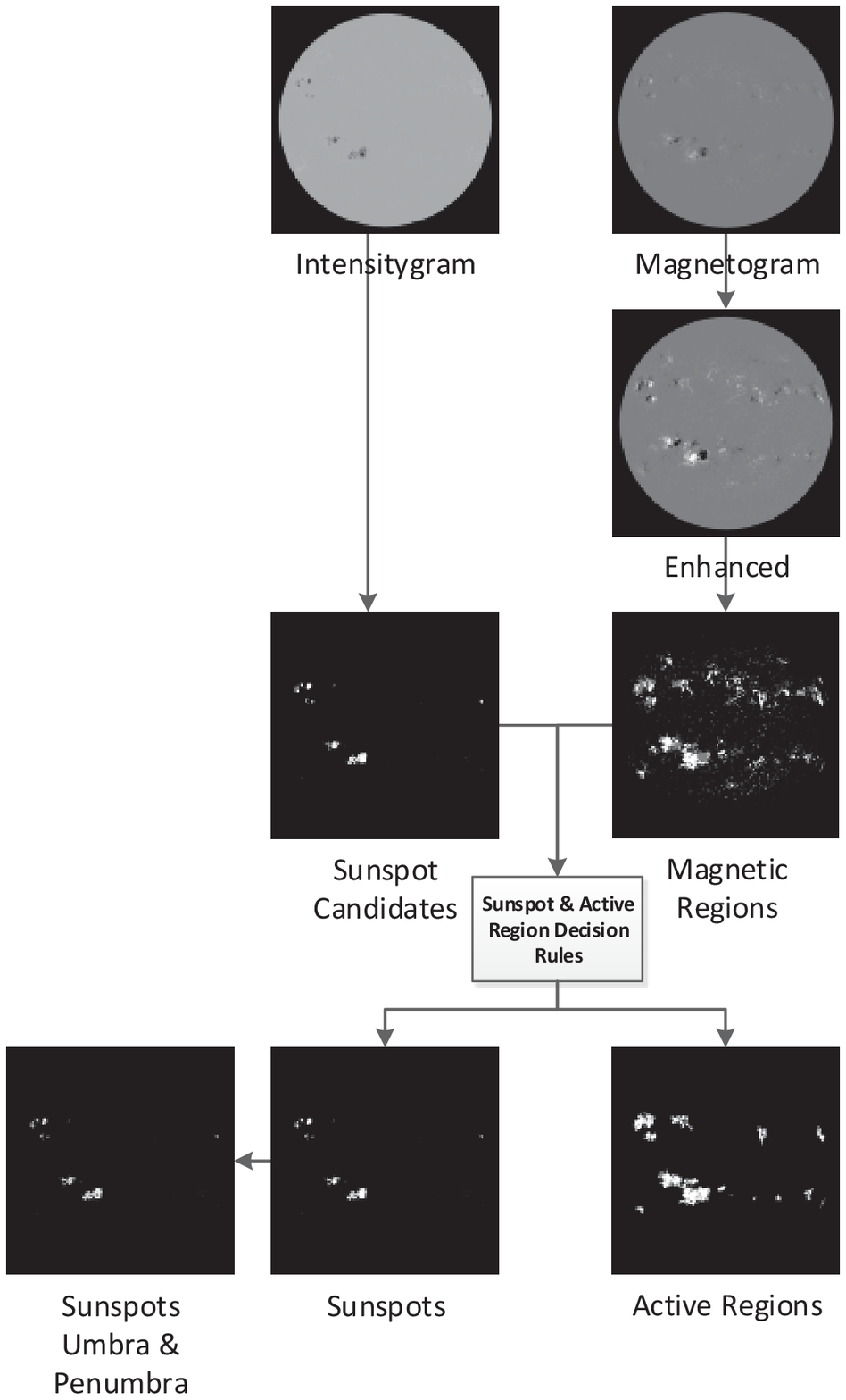}
	\caption{\label{fig:19}
		Flowchart of the individual steps of the feature identification scheme of ASAP.}
\end{figure*}

This section described the proposed approach for detecting a number of solar features, present in MDI intensitygrams and magnetograms. The feature detection criteria are summarised in Table\,\ref{T:Features}, while the overall detection process is summarised as a flowchart and shown in Figure\ref{fig:19}. The key elements of the feature identification scheme are as follows:
\begin{itemize}
\item The detection of solar features present in MDI intensitygrams and magnetograms, in line with ASAP feature detection.
\item The detection of faculae and network is based on the location of magnetic regions in relation to active regions.
\item The automated detection and classification of solar features and the calculation of their area coverage for the whole period of MDI data. 

A summary of the solar features detection processes is given below:
\item Detect initial sunspots candidates from intensitygrams, and magnetic regions from magnetograms. Un-magnetised regions on the magnetogram define the quiet Sun.
\item Group sunspots and magnetic regions for the detection and verification of active regions and sunspots. Active regions are detected based on the localisation of magnetic regions and sunspots using NNs.
\item Identify faculae and network by subtracting sunspot map from the magnetic region map. The remaining magnetic regions pixels are classified as faculae if they are located within active regions, otherwise, if they are located outside active regions, then they are classified as network.  
\end{itemize}

\begin{table}
	\caption{The solar features identification criteria that have been adopted in this work, and the parameters employed to identify them, from the respective data set. Intensitygrams and Magnetograms are shortened to Int and Mag, respectively.}
	\begin{center}
		{\scriptsize 
			\begin{tabular}{llll}
				\hline
				\noalign{\smallskip}
				\noalign{\smallskip}
				Feature & \parbox[t]{2cm}{Data type} & Detection Criterium &  Description \\
				\noalign{\smallskip}
				\noalign{\smallskip}
				\hline
				\noalign{\smallskip}
				\noalign{\smallskip}
Sunspots  & Int  &  \parbox[t]{4cm}{$Th_{\rm ss}=\left\langle I \right\rangle - f_{\rm ss}$} & \parbox[t]{7cm}{where $f_{\rm ss}$=15 } \\
\noalign{\smallskip}
\noalign{\smallskip}				
Pores	& Int  	& \parbox[t]{4cm}{$Th_{\rm pores}=\left\langle I \right\rangle_{\rm region} - \left( f_{\rm p} \times \sigma_{\rm I,SMin} \right) $} & \parbox[t]{7cm}{where $f_{\rm p}$= 7, and $\sigma_{\rm I,SMin}$=1.5} \\	
		\noalign{\smallskip}
		\noalign{\smallskip}
				
\parbox[t]{2cm}{Umbra, Penumbra} & Int   &  	\parbox[t]{4cm}{$Th_{\rm up}=\left\langle I \right\rangle_{\rm ss} - \sigma_{\rm ss}  $}	&    \parbox[t]{7cm}{where $\left\langle I \right\rangle_{\rm ss}$ is the mean of sunspot groups and $\sigma_{\rm ss}$ is the standard deviation of sunspot group. Umbra and penumbra are defined after the detection of sunspots. Sunspot pixels with values below the threshold are classified as umbra, while the remaining sunspot pixels are classified as penumbra.}\\					
\noalign{\smallskip}
\noalign{\smallskip}
				
Faculae	& Mag  & \parbox[t]{4cm}{$Th_{\rm mr}= \left\langle B \right\rangle \pm \left( f_{\rm mr} \times \sigma_{\rm B,SMin}\right)$} & \parbox[t]{7cm}{where $f_{\rm mr}$ = 2 and $\sigma_{\rm B,SMin}$ = 2.4.} 	 \\

 	    & & & \parbox[t]{7cm}{Faculae are magnetic regions, except sunspots, located within active regions boundaries.} 	 \\
 				
\noalign{\smallskip}
\noalign{\smallskip}								
				
Network 	& Mag   & \parbox[t]{4cm}{$Th_{\rm mr}= \left\langle B \right\rangle \pm \left( f_{\rm mr} \times \sigma_{\rm B,SMin}\right)$} 	& \parbox[t]{7cm}{where $f_{\rm mr}$ = 2 and $\sigma_{\rm B,SMin}$ = 2.4. }\\

 	       & & & \parbox[t]{7cm}{Network are magnetic regions located outside active regions.}\\

				\noalign{\smallskip}
				\hline
			\end{tabular}
			\label{T:Features}}
	\end{center}
\end{table}

\section{Results}\label{sec:results}
\begin{figure*}
	\centering
	\ Solar Max., May 17, 2000 01:39 UT \\
	\includegraphics[width=0.45\textwidth]{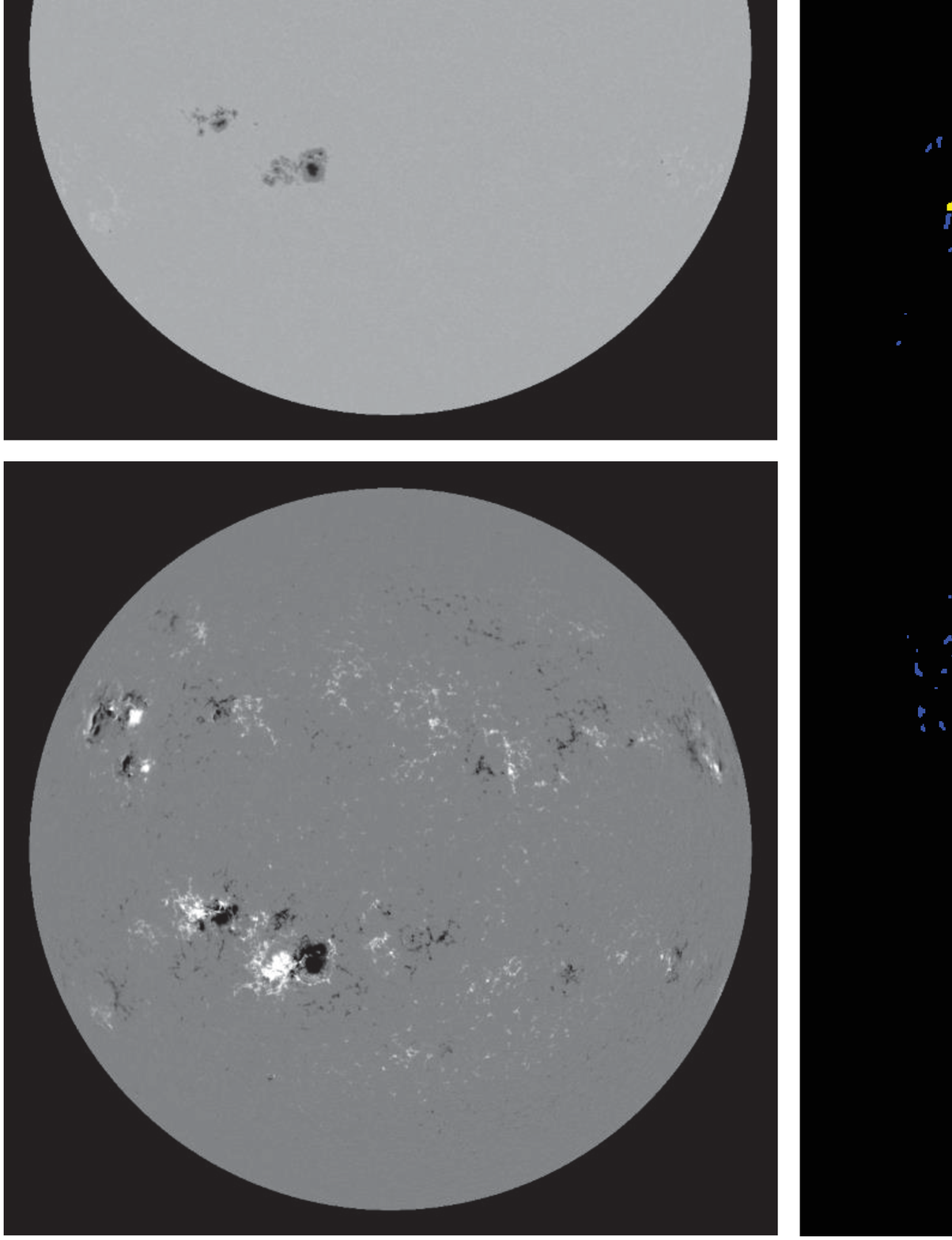}\\
	\ Solar Min., April 22, 1997 00:04 UT  \\ 
	\includegraphics[width=0.45\textwidth]{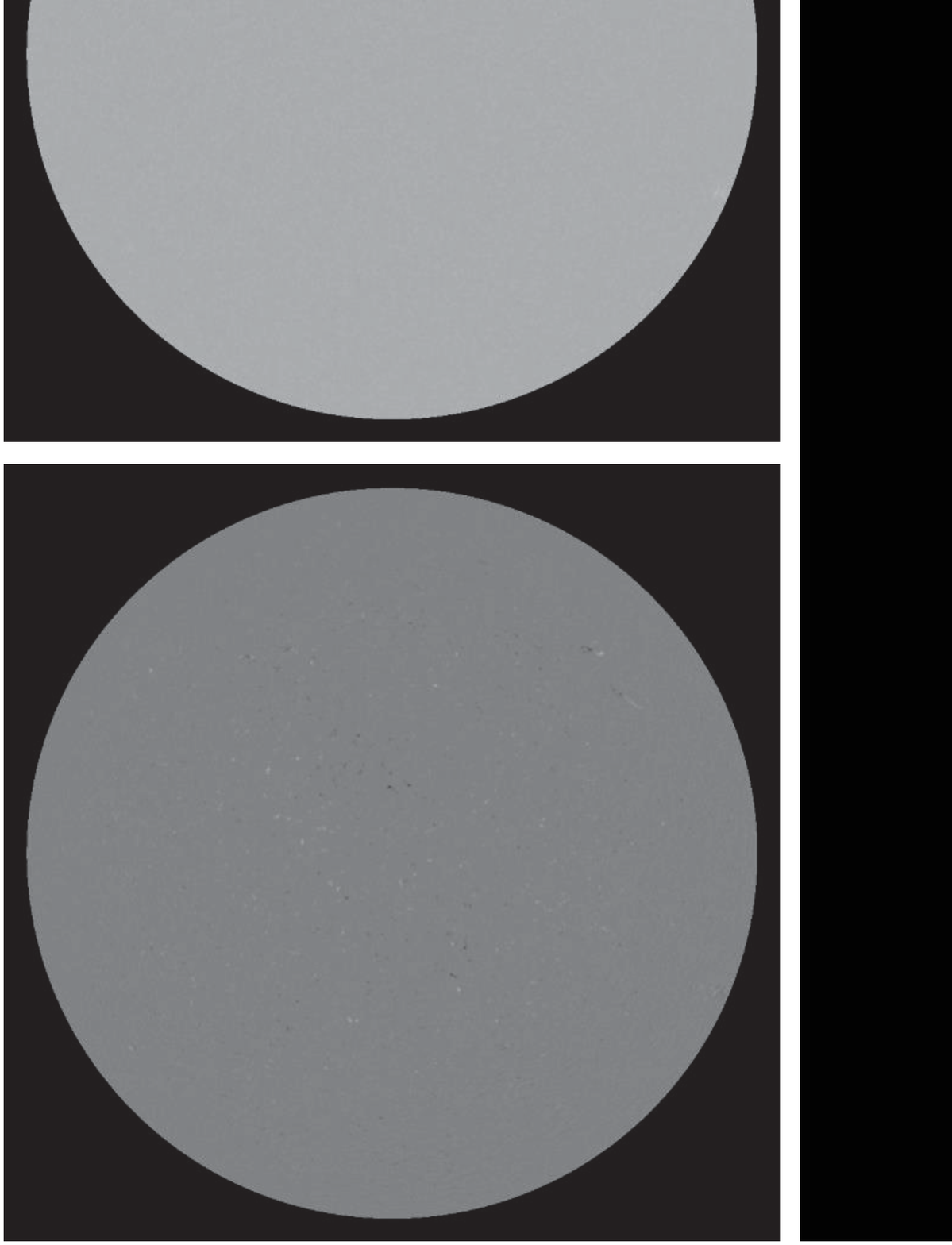}
		\caption{\label{fig:result}
		The left panels shows the original MDI intensitygram (top) and magnetogram (bottom). The right panels shows a colour map of the detected solar features: umbra (red), penumbra (orange), faculae (yellow), and network (blue).}
\end{figure*}

The main aim of this research is the detection of solar features for the provision of useful proxies for reconstructing TSI and SSI. For the detection of photospheric features, we carried out a series of processes to identify feature of interest on MDI intensitygrams and magnetograms. Figure\,\ref{fig:result} shows colour maps of features extracted from intensitygram and magnetogram pairs taken from solar maxima and the solar minima on the days indicated in the figure.
Further to the application of the identified solar features, we have also computed the area filling factors for the detected features. These will be used in our future work for the SSI reconstruction. The calculated filling factors for a feature reflect the fraction of the solar disk covered by the feature and assign reference synthetic spectra to it. The filling factor is specified as a function of radial position on the solar disk, by relative radius values that determine concentric rings around the disk centre, as indicated in Tables\,\ref{T:ff1} and \ref{T:ff2}. In the case of MDI the last interval would be between 0.95 and the limb. To reduce uncertainty in the calculation of area filling factors, the areas are calculated to sub-pixel accuracy. Tables\,\ref{T:ff1} and \ref{T:ff2} give the sets of filling factors calculated for the intensitygram-magnetogram pairs for May 17, 2000 11:11 UT and April 22, 1997 at 00:04 UT respectively. The bottom row in each table is provided as a check and shows the number of pixels for each type of feature and should equal the sum of the numbers in the rows above.

\begin{table}[h!]
	\caption{Pixel coverage of the rings with inner and outer radii, $R_1$ and $R_2$ respectively, expressed relative to the solar radius, of the features penumbra, umbra, faculae, and network, as detected on the intensitygram and the magnetogram observed May 17, 2000, 01:39 UT.}
	\begin{center}
		{\scriptsize
			\begin{tabular}{rrrrrrrrr}
				\hline
				\noalign{\smallskip}
				Index &	$R_1$ &	$R_2$ &	Penumbra &	Umbra & Faculae & Network \\					
				\noalign{\smallskip}				
				\hline
				\noalign{\smallskip}			
				1&	0.00&	0.07&	0.00&		0.00&	0.00&		57.47     \\
				2&	0.07&	0.16&	0.00&		0.00&	0.00&	 	390.68    \\
				3&	0.16&	0.25&	1.00&		0.00&	135.42&		1,275.05  \\
				4&	0.25&	0.35&	311.79&		17.00&	1,249.22&	2,871.43  \\
				5&	0.35&	0.45&	1,639.00&	274.38&	4,140.83&	3,648.80  \\
				6&	0.45&	0.55&	1122.68&	113.61&	4,658.81&	4,720.42  \\
				7&	0.55&	0.65&	263.51&		10.00&	3,665.51&	4,158.12  \\
				8&	0.65&	0.75&	126.00&		5.00&	709.90&		4,462.60  \\
				9&	0.75&	0.85&	580.73&		100.00&	2,635.51& 	3,148.56  \\
				10&	0.85&	0.95&	497.26&		57.00&	2,993.76&	2,713.37  \\
				11&	0.95&	1.00&	6.00&		0.00&	149.00&		478.44	  \\	
			\multicolumn{3}{c}{Sum}& 4548.00&  	577.00& 20,338.00&  27,925.00 \\
				\noalign{\smallskip}
				\hline
			\end{tabular}
			\label{T:ff1}}

	\caption{Same as Table\,\ref{T:ff1} but for the intensitygram and the magnetogram observed April 22, 1997, 00:04 UT.}
		{\scriptsize 
			\begin{tabular}{rrrrrrrrr}
				\hline
				\noalign{\smallskip}
				Index &	$R_1$ &	$R_2$ &	Penumbra & Umbra & Faculae & Network	\\
				\noalign{\smallskip}				
				\hline
				\noalign{\smallskip}			
				1 &	0.00&	0.07&	0.00&	0.00&	0.00&	26.97	\\
				2&	0.07&	0.16&   0.00&	0.00&	0.00&	144.31	\\
				3&	0.16&	0.25&	0.00&	0.00&	0.00&	479.62	\\
				4&	0.25&	0.35&	0.00&	0.00&	0.00&	499.07 	\\
				5&	0.35&	0.45&	0.00&	0.00&	0.00&	571.25	\\
				6&	0.45&	0.55&	0.00&	0.00&	0.00&	542.30	\\
				7&	0.55&	0.65&	0.00&	0.00&	0.00&	537.65	\\
				8&	0.65&	0.75&	0.00&	0.00&	0.00&	395.57	\\
				9&	0.75&	0.85&	0.00&	0.00&	0.00&	404.29	\\
				10&	0.85&	0.95&	0.00&	0.00&	0.00&	84.66 	\\
				11&	0.95&	1.00&	0.00&	0.00&	0.00&	27.25	\\	
				\multicolumn{3}{c}{Sum} & 0.00&	0.00&	0.00&	3,713.00\\
				\noalign{\smallskip}
				\hline
			\end{tabular}
			\label{T:ff2}}
	\end{center}
\end{table}
\begin{figure*}
	\centering
	\includegraphics[width=.45\textwidth]{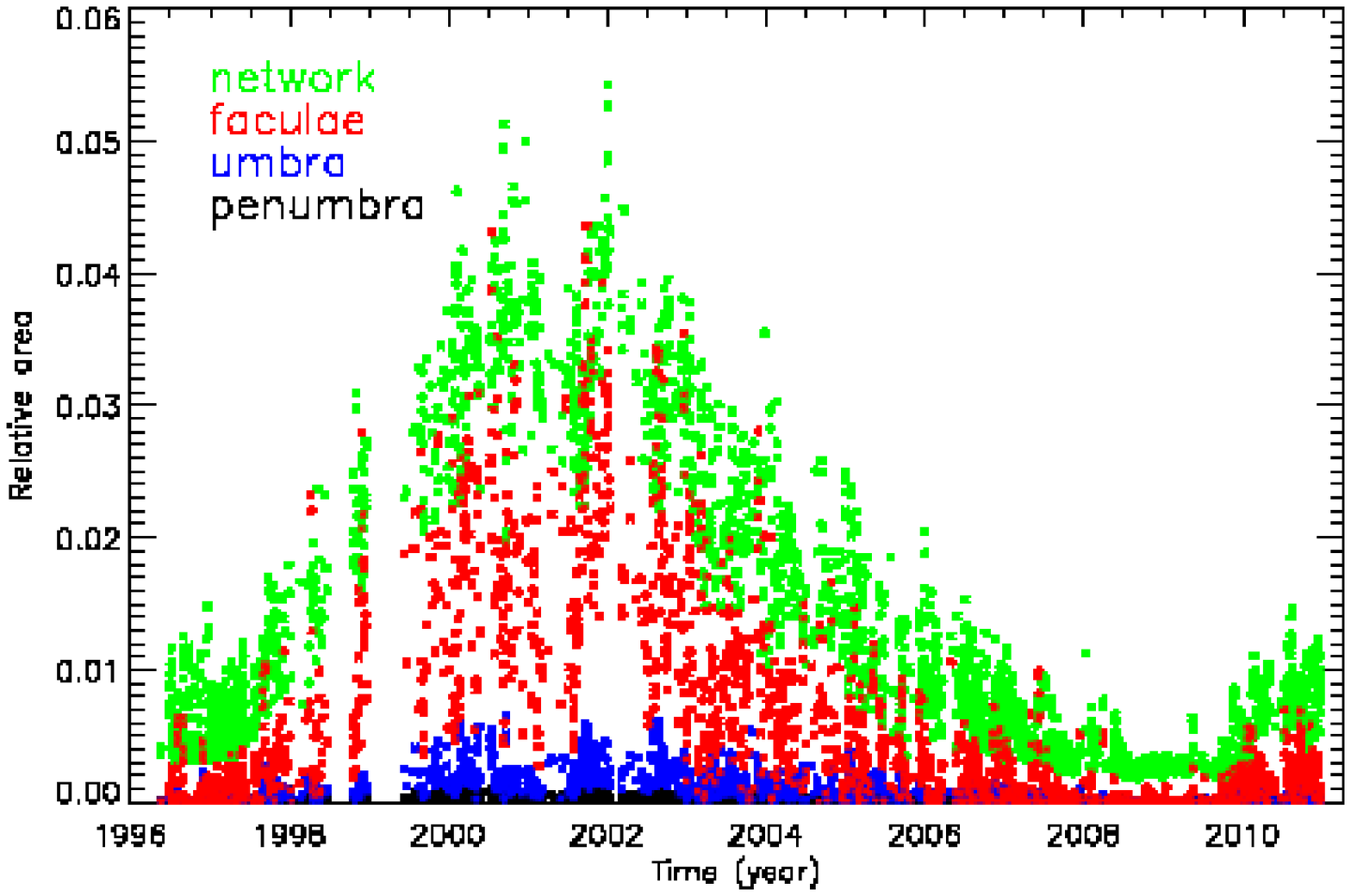}
	\includegraphics[width=.45\textwidth]{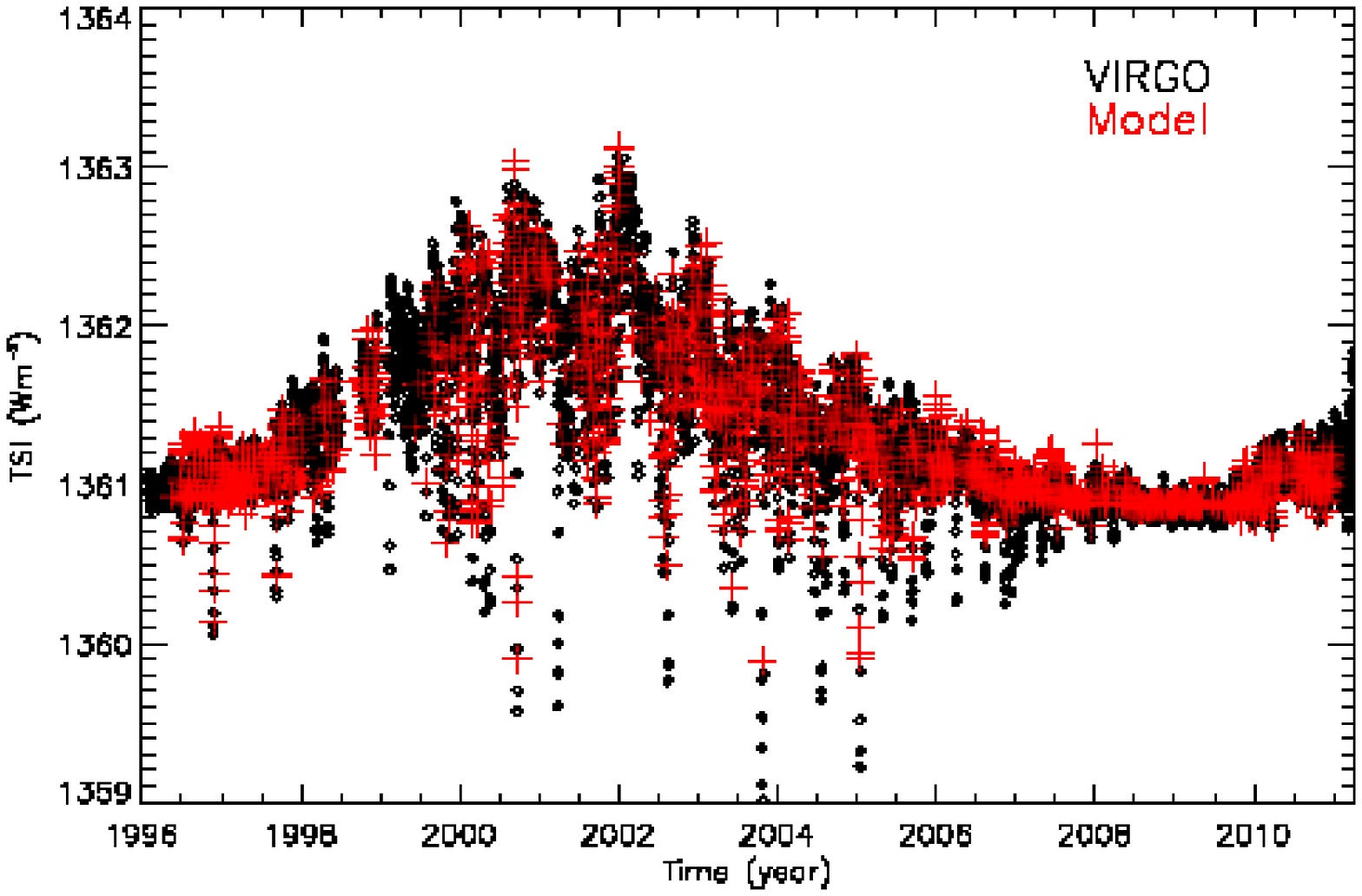}
	\caption{\label{fig:21}
		The left panel gives the relative area of the full solar disk of network (green), faculae (red), umbra (blue), and penumbra (black). The right panel compares the fitted TSI data (red) with the VIRGO data corrected to the SI-scale (black).}
\end{figure*}

\begin{figure*}
	\centering
	\includegraphics[width=.45\textwidth]{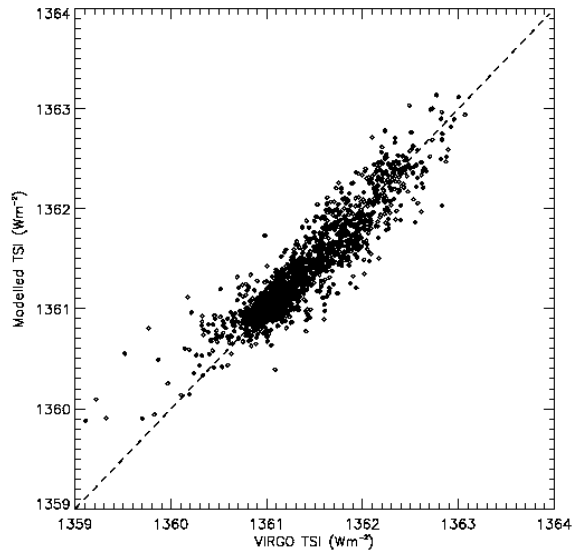}
	\caption{\label{fig:22}
		Scatter plot between the VIRGO TSI data and the fitted data based on our regression analysis. The squared correlation coefficients of the fit versus the VIRGO observation is $r^2=0.88$. The dashed line indicates equality between the two data sets.}
\end{figure*}

The area coverage of the identified features (umbra, penumbra, network and faculae), which will be employed in upcoming reconstructions, is given in Figure\,\ref{fig:21} (left panel). The solar cycle modulation of the area coverage of all the features is clearly visible. To further test the segmentation result, we carried out a multi-regression fit to the SOHO/VIRGO data \citep{Froehlich1997}. Following \cite{Fehlmann2012} we first corrected (divided) the VIRGO data by the correction factor 1.0034 to account for the difference between the World Radiometric Reference (WRR) maintained at PMOD/WRC to which VIRGO has been compared to, and the SI-scale, the latter being realized at the TSI Radiometer Facility \citep[TRF,][]{Kopp2012}. The Picard/PREMOS instrument \citep{PREMOS,Schmutz2013} has been compared to both scales and independently confirmed the difference of the WRR and SI scale. We then carried out the multi-linear regression from which we obtained correlation coefficients between the observed TSI time series and the disk-integrated relative area coverage of umbra, penumbra, network, and faculae of 0.29, 0.29, 0.82, and 0.57, respectively.
Figure\,\ref{fig:21} (right panel) shows the comparison of the model (red) with the VIRGO data (black). It is clear that the TSI variability is very well pre-produced. Some strong decrease in TSI due to the appearance of large sunspots is slightly underestimated. Figure\,\ref{fig:22} shows the scatter plot between the VIRGO data and the fitted data, the correlation coefficient between the two being $r^2=0.88$. 

Figure\,\ref{fig:22} shows the scatter plot of the VIRGO data compared to the simple regression model. The analysis of the time series shows that the applied method to identify solar activity features from SOHO/MDI intensitygrams and magnetograms gives realistic results.
\section{Conclusions}\label{sec:concl}
In this paper, a new image segmentation approach is introduced to enable the efficient detection of sunspot umbra and penumbra, faculae, network and quiet Sun. Other techniques were also developed for the visualisation and the calculation of area coverages of the detected features. The imaging methods and tools introduced in this work enabled the extraction of area coverages for the different solar features for the duration of the entire MDI mission. These area coverages will be used in the near future for the reconstruction of SSI.
We extended the ASAP system for the purpose of identifying a number of solar features and calculating their area coverage, which is necessary for the SSI reconstruction. As part of this process, we carried out investigations to determine a set of key values that enables thresholding of solar features on MDI intensitygrams and magnetograms. ASAP is an automated system that provides near real-time detection of solar features and prediction of solar flares, and is currently operating using SDO HMI intensitygrams and magnetograms. Taking this into account, the approaches described in this paper can be applied to a calibrated SDO FITS files in order to enable the systematic detection of solar features, and subsequently enable the provision of near real-time solar irradiance using SDO HMI images. 

The relative contribution of the quiet Sun and various solar activity features to the total solar irradiance strongly depends on wavelength. Radiometers, e.g., PICARD/PREMOS \citep{Schmutz2009} and PROBA2/LYRA \citep{LYRA2006}, {dedicated to observe} the solar spectrum in specific passbands from the EUV to the visible and near infrared, which allows a detailed validation of our approach. The situation regarding SSI measurements is however rather bleak. The only upcoming SSI instrument currently planned for the future is the TSIS instrument which will be integrated on the international space station in 2017. Therefore, it is crucial to have reliable models available that allow modelling of the expected SSI in a robust way. Using the results presented here along with the synthetic spectra calculated from semi-empirical models will allow us to determine the contributions of the various features to the variation in SSI. As the intensity spectrum emitted by the different solar components can be pre-calculated, the solar spectrum can be predicted practically in real-time. The reconstruction of the SSI over a broad wavelength range will be the focus of a follow-up paper.
\begin{acknowledgements}
SOHO is a project of international cooperation between ESA and NASA. The authors acknowledge that the research leading to these results has received funding from the European Community's Seventh Framework Programme (FP7 2012) under grant agreement no 313188 (SOLID). The authors thank the SOI and MDI team for their guidance to acquire MDI data, and thank Dr. Mohammad Alomari for his assistance in retrieving MDI data. The authors thank the editor and the anonymous reviewers for their careful revision and valuable comments. The editor thanks two anonymous referees for their assistance in evaluating this paper.
\end{acknowledgements}
\bibliographystyle{swsc}
\bibliography{ref_all11_swsc,biblio_spoca}
\end{document}